\documentclass[11pt]{article}
\usepackage{axodraw}
\usepackage{amsmath,amsfonts,sint,epsfig,cite,graphics}
\usepackage{mathrsfs}
\usepackage{multirow}
\usepackage{macros_static}
\usepackage{wrapfig}
\usepackage{color}
\usepackage{soul}
\usepackage{colortbl}
\usepackage{mathbbol}
\usepackage{rotating}
\usepackage{bm}

\usepackage[english]{babel}

\usepackage{psfrag}
\usepackage{amsmath}
\usepackage{subfigure}
\usepackage{feynmp}
\usepackage{amsfonts}

\newcommand{\bdm}{\begin{displaymath}}
\newcommand{\edm}{\end{displaymath}}
\renewcommand{\be}{\begin{equation}}
\renewcommand{\ee}{\end{equation}}
\renewcommand{\bi}{\begin{itemize}}
\renewcommand{\ei}{\end{itemize}}

\newcommand{\fpipi}{{f_{\pi\pi}(Q^2)}}
\newcommand{\fpipizero}{{f_{\pi\pi}(0)}}

\newcommand{\refeq}[1]{\eqref{#1}}

\newcommand{\Order}[1]{\ensuremath{\mathcal{O}(#1)}}
\newcommand{\Ord}[1]{\Order{#1}}

\newcommand{\fpp}{f_{\pi\pi}}
\newcommand{\xt}{x_2}
\newcommand{\hxt}{\hat{x}_2}
\newcommand{\bq}{\bar{q}^2}
\newcommand{\bqf}{\bar{q}^4}
\newcommand{\hbq}{\hat{q}^2}

\newcommand{\lc}[1]{\bar{\ell}_{#1}}
\newcommand{\lcr}[1]{\ell_{#1}^r}
\newcommand{\hlcr}[1]{\hat{\ell}_{#1}^r}
\newcommand{\ltil}{\tilde{\ell}}

\newcommand{\rmr}{r_m^r}
\newcommand{\rfr}{r_f^r}
\newcommand{\rvrt}{r_{V2}^r}
\newcommand{\rvro}{r_{V1}^r}
\newcommand{\mnlo}{\left[\mpi\right]_1}
\newcommand{\mnnlo}{\left[\mpi\right]_2}
\newcommand{\fnlo}{\left[\fpi\right]_1}
\newcommand{\fnnlo}{\left[\fpi\right]_2}
\newcommand{\ffnlo}{\left[\fpp\right]_1}
\newcommand{\rpi}{\left\langle r_\pi^2 \right\rangle}
\newcommand{\pvec}{\vec p}
\newcommand{\ubar}{\bar u}
\newcommand{\dbar}{\bar d}
\newcommand{\rsq}{\langle r^2_\pi\rangle}

\newcommand{\mbf}[1]{\mbox{\boldmath${#1}$}}
\newcommand{\mbfsm}[1]{\mbox{\scriptsize\boldmath${#1}$}}
\def\pvec{\mbf{p}}
\def\nvec{\mbf{n}}
\def\xvec{\mbf{x}}
\def\pvecsm{\mbfsm{p}}

\def\xvecsm{\mbfsm{x}}
\def\thvecsm{\mbfsm{\theta}}

\include{include}

\begin{document}

\newcommand{\PH}{P_{\rm H}}
\newcommand{\PL}{P_{\rm L}}
\newcommand{\unity}{{I}}

\def\slash#1{\mkern-1.5mu\raise0.6pt\hbox{$\not$}\mkern1.2mu #1\mkern 0.7mu}

\begin{titlepage}

\begin{flushright}
\vskip .4cm
HIM-2013-03\\
MITP/13-029
\end{flushright}

\vskip 0.5cm
\begin{center}
{\Large\bf 
The pion vector form factor from lattice QCD\\[0.3cm]
and NNLO chiral perturbation theory}
\end{center}
\vskip 1.0cm
\begin{center}
{\large
   Bastian~B.~Brandt$^{a}$,
   Andreas~J\"uttner$^{b}$ and
   Hartmut~Wittig$^{c,d}$ 
}
\vskip 0.5cm
$^{a}$\,Institut~f\"ur~theoretische~Physik, University~of~Regensburg, 
 D-93040~Regensburg, Germany
\vskip 0.3cm
$^{b}$\,School of Physics and Astronomy, University of Southampton,
  Southampton SO17~1BJ, United Kingdom
\vskip 0.3cm
$^{c}$\,PRISMA Cluster of Excellence, Institut~f\"ur~Kernphysik,
 University of Mainz, D-55099 Mainz, Germany
\vskip 0.3cm
$^{d}$\,Helmholtz Institute Mainz, University of Mainz, D-55099 Mainz,
  Germany 
\vskip .5cm
{\bf Abstract}
\vskip 0.5ex
\end{center}

\noindent

We present a comprehensive study of the electromagnetic form factor, the decay constant and the mass of the
pion computed in lattice QCD with two degenerate O($a$)-improved Wilson quarks
at three different lattice spacings in the range $0.05-0.08$\,fm and pion
masses between 280 and 630~MeV at $m_\pi\,L\geq4$. Using partially twisted
boundary conditions and stochastic estimators, we obtain a dense set of
precise data points for the form factor at very small momentum transfers,
allowing for a model-independent extraction of the charge radius. 
Chiral Perturbation Theory (ChPT) augmented by terms which model lattice
artefacts {is then compared to the data}. 
{At next-to-leading order the effective theory
fails to produce a consistent 
description of the full set of pion observables but describes the data well
when only the decay constant and mass are considered.}
By contrast, using the next-to-next-to-leading order
expressions to perform global fits
result in a consistent description of all data. We obtain
$\rpi=0.481(33)(13)\, \fm^2$ as our final result for the
charge radius at the physical point. Our calculation also yields estimates for
the pion decay constant in the chiral limit, $F_\pi/F=1.080(16)(6)$, the quark
condensate, $\Sigma^{1/3}_{\msbar}(2\,\GeV)=261(13)(1)\,\MeV$ and several
low-energy constants of SU(2) ChPT.
\vskip 1cm
\noindent{\it Keywords:}
Chiral Perturbation Theory; Lattice QCD;
\vskip 0.0ex

\vfill
\eject

\end{titlepage}

\tableofcontents
\section{Introduction}\label{sec:introduction}

Thanks to the continued progress in improving numerical and field theoretical
techniques~\cite{Hasenbusch:2001ne,Luscher:2005rx,Urbach:2005ji,Clark:2006fx,Luscher:2007se,Luscher:2007es,Luscher:2008tw,Marinkovic:2010eg},
computer simulations of QCD on a Euclidean space-time lattice are sufficiently
advanced to produce reliable results for a number of phenomenologically
important quantities (see for example the FLAG-summary
\cite{Colangelo:2010et}). Some of these results are postdictions which can
serve as a test of lattice QCD as a tool, other results are real predictions
which can be used to address the validity of the Standard Model. In both cases
the estimation of systematic uncertainties is a crucial but often delicate
issue.

In the case at hand, i.e. the pion electromagnetic form factor, the dominant
systematic in recent
calculations~\cite{Brommel:2006ww,Frezzotti:2008dr,Boyle:2008yd,Aoki:2009qn,Nguyen:2011ek,Brandt:2011jk}
is due to its strong quark-mass dependence which complicates the extrapolation
from unphysically heavy quark masses to the physical point. Chiral
perturbation theory can provide guidance here. The corresponding expressions
for the form factor as a function of the quark mass have been worked out up to
NNLO~\cite{Gasser:1984gg,Gasser:1984ux,Bijnens:1998fm,Bijnens:2002hp}.
However, concerning the effective theory's validity, a particular concern here
is the tree-level contribution of vector degrees of freedom which can couple
to the probing photon. In the effective theory these have been
\textit{integrated out} and enter only passively through the low-energy
parameters in the effective Lagrangian. The scale separation between the
Goldstone bosons ($\pi$, K, $\eta$) and the vector bosons ($\rho$, $\omega$)
is, however, not large, and one may be worried about the applicability of the
low-energy effective theory. Other interesting observables like the hadronic
vacuum polarisation do also receive tree-level contributions from vector
particles, and similar concerns can be
raised~\cite{Aubin:2006xv,Feng:2011zk,Boyle:2011hu,DellaMorte:2010aq,DellaMorte:2011aa}.

From a lattice practitioners point of view the mere evaluation of the pion
form factor on a given lattice ensemble is a rather straightforward task, and
therefore this quantity serves as an ideal laboratory for studying the above
questions.
Our strategy is to compare lattice QCD results to the predictions of chiral
effective theory for the pion form factor and charge radius, the pion decay
constant and its mass. For the pion mass and decay constant the expressions of
ChPT are known to provide a good description of lattice data in the range of
quark masses studied in this work (see~\cite{Colangelo:2010et}). In the same
spirit we have concentrated on producing data for the form factor for very
small space-like photon momenta, in order to remain in the realm of chiral
perturbation theory. To this end we made extensive use of partially twisted
fermionic boundary
conditions~\cite{Bedaque:2004kc,deDivitiis:2004kq,Sachrajda:2004mi,Boyle:2007wg},
which allowed us to induce small pion momenta despite simulating in a finite
lattice volume. Thereby we were able to determine the pion charge radius in a
quasi model-independent way. By comparing a variety of fit {\it ans\"atze}
based on Chiral Perturbation Theory (ChPT) at NLO and NNLO respectively, we
investigated whether different fits provide a consistent description of the
data and lead to reliable results for the pion charge radius and decay
constant. For our final estimates we have performed an elaborate analysis of
systematic uncertainties arising from lattice artefacts and finite-volume
effects. As a byproduct we have determined the relevant low-energy constants
(LECs) of two-flavour QCD at NNLO.

We briefly anticipate the core results: On a qualitative level we note that a
{\it joint} description of our data for the pion mass, decay constant and form
factor in terms of ChPT at NLO fails, while a consistent description of all
three quantities in terms of ChPT can only be achieved at NNLO. We stress that
the validity of this statement must be monitored as the pion mass is further
decreased. In fact, our findings emphasise the importance of performing
simulations at or very near the physical point.

Our final results for decay constants, LECs and the charge radius are
\begin{equation}
\begin{array}{llllllll}
F_\pi  & = & 90(8)(2)\,{\rm MeV}\,, & F     & = & 84(8)(2)\,{\rm MeV}\,,\\
F_\pi/F& = & 1.080(16)(6)\,,        & \Sigma^{1/3} & = & 261 \:(13)(1)\,{\rm MeV}\,,&{\rm from\, NLO\, fit}\,,\\
\lc{3} & = & 3.0(7)(5)\,,           &\lc{4} & = & 4.7(4)(1)\,,
\\[4mm]
\rpi   & = & 0.481(33)(13)\,{\fm}^2,&&&&\multirow{2}{*}{\rm from\, NNLO\, fit\,,}\\
\lc{6} & = & 15.5(1.7)(1.3),\\
\end{array}
\end{equation}
where the quark condensate $\Sigma$ is defined in the $\msbar$-scheme at a
renormalisation scale of 2~GeV.

The outline of this paper is as follows: In section~\ref{sec:compstrat} we
introduce the basic definitions and our computational setup. Simulation
details and lattice results are presented in section~\ref{sec:Latticesetup},
followed by their discussion in terms of fits and extrapolations in
section~\ref{sec:extrapolations}. Our conclusions are presented in
section~\ref{sec:conclusions}. Preliminary reports of the results included in
this paper have appeared
in~\cite{Brandt:2010ed,Brandt:2011sj,Brandt:2011jk,Brandt:2013mb}.

\section{Computational strategy}\label{sec:compstrat}
In this section we define the pion decay constant $F_\pi$, the 
pion mass $m_\pi$ and the pion charge radius $\rpi$
in terms of
Euclidean two- and three-point functions.

The electromagnetic form factor in two-flavour QCD is defined by
\be
\label{eq:fpipi}
   \left\langle\pi^+(\pvec_f)|
   {\textstyle\frac{2}{3}}\ubar\gamma_\mu u
  -{\textstyle\frac{1}{3}}\dbar\gamma_\mu d
   |\pi^+(\pvec_i)\right\rangle = (p_f+p_i)_\mu\,f_{\pi\pi}(q^2)\,,
\ee
where $q^2=(p_f-p_i)^2$ is the space-like momentum transfer,
$-q^2\equiv Q^2\geq0$. Near vanishing momentum transfer, the form
factor can be expanded in powers of $q^2$. By convention, the linear
term defines the charge radius, $\langle r^2_\pi\rangle$, i.e.
\be
\label{eq:chrad}
       f_{\pi\pi}(q^2)=1-\frac{1}{6}\langle r_\pi^2\rangle
       q^2+\rmO(q^4), \qquad
       \langle r^2_\pi\rangle = 6\left.
       \frac{\rmd f_{\pi\pi}(q^2)}{\rmd q^2}\right|_{q^2=0}.
\ee
Simulations of lattice QCD are necessarily performed in a finite
volume, and hence the accessible range of momentum transfers is rather
limited. In a conventional setting (periodic fermionic boundary conditions)
the initial and final pions can
only assume the Fourier momenta, $\nvec\,2\pi/L$, where $\nvec$ is a
vector of integers. Unless one can afford to simulate very large box
sizes $L$, the lowest non-zero value of $Q^2$ can be rather
sizeable. It is then doubtful whether the charge radius can be
determined in a model-independent fashion, e.g. from the linear slope
of the form factor near vanishing $Q^2$.

Partially flavour-twisted boundary conditions
\cite{Bedaque:2004kc,Sachrajda:2004mi,Flynn:2005in,deDivitiis:2004kq,Boyle:2007wg}
have by now become a standard tool to overcome this problem. By
imposing periodicity on the quark fields in the spatial directions up
to a phase factor, i.e.
\be
  \psi(x+\mbf{\hat e}_j L)=\psi(x)e^{i\theta_j/L}\,\quad j=1,2,3,
  \label{eq:twisted_bc}
\ee
it was shown in~\cite{Boyle:2008yd} that the momentum transfer
satisfies
\be
\label{eq:qsq}
    -Q^2\equiv q^2=(p_f-p_i)^2 =
    \Big[E_\pi\Big(\pvec_f{+\frac{\mbf\theta_f}L}\Big)
    -E_\pi\Big(\pvec_i{+\frac{\mbf\theta_i}L}\Big)\Big]^2
  -\Big[\Big(\pvec_f+\frac{\mbf\theta_f}{L}\Big)
  -\Big(\pvec_i+\frac{\mbf\theta_i}{L}\Big) \Big]^2\,.
\ee
Here, $\mbf{\theta}_i$ and $\mbf{\theta}_f$ denote the vectors of
twist angles applied to the quark probed by the electromagnetic
current in the initial and final pion, respectively, and 
{
\be
\label{eq:dispersion}
E_\pi(\pvec)= \sqrt{m_\pi^2+\pvec^2}\,,
\ee
is the pion dispersion relation~\cite{Flynn:2005in}}. In this work we
have paid particular attention to choosing twist angles which result
in a very dense set of data points in the immediate vicinity of
$Q^2=0$, such that the charge radius could be determined by means of a
discretised derivative of the form factor.

\subsection{Euclidean correlation functions}

All our calculations have been performed in two-flavour QCD, employing
$\rmO(a)$ improved Wilson fermions~\cite{Wilson:1974sk}. 
We have used the non-perturbative
determination \cite{Jansen:1998mx} of the improvement coefficient
$\csw$ which multiplies the Sheikholeslami-Wohlert term.

In this work we consider correlation functions of the non-singlet,
$\rmO(a)$ improved axial current and pseudoscalar density
\cite{Luscher:1996sc},
\bea
    A^I_\mu(x)&=&\bar u(x) \gamma_\mu\gamma_5 d(x)
     +a\ca\tilde\partial_\mu P(x)\,, \\
    P(x)&=&\bar u(x) \gamma_5 d(x),
\eea
as well as the $\rmO(a)$ improved electromagnetic current, i.e.
\be
 V^I_\mu(x) = V_\mu(x) + a\cv\tilde\partial_\nu T_{\mu\nu}(x)\,.
     \label{eq:Vimpr2}
\ee
{Since we are simulating mass-degenerate light quarks it is 
sufficient to consider only the local vector and tensor currents
$V_\mu(x)=\bar d(x)\gamma_5 u(x)$ and 
$T_{\mu\nu}(x)=i \bar q(x)\sigma_{\mu\nu}q(x)$, respectively.}
In the above expressions,
$\tilde \partial_\mu$ is the symmetrised discrete derivative in
direction $\mu$. The improvement coefficient $\ca\equiv\ca(g_0)$ has
been computed non-perturbatively in two-flavour QCD
in~\cite{DellaMorte:2005se} for a range of bare couplings~$g_0$ (see also
table~\ref{tab:coeffs}) and 
$\cv$ has been computed in one-loop perturbation theory~\cite{Sint:1997dj} 
but at the end of this section we will argue that it is of little relevance 
here. Note that the hadronic
matrix elements of the currents must still be renormalised in order to
yield physical observables.

We compute the following two-point functions
\bea
  & & C_{PP}(t,\pvec) =
  \sum\limits_{\xvecsm}e^{i{\pvecsm\cdot\xvecsm}} \langle
  \,P(t,\xvec)\, P^\dagger(0,\mbf{0})\,\rangle \simeq
  \frac{|\mathcal{Z}_P|^2}{2E(\pvec)} \left( 
     e^{-E({\pvecsm})\,t} + e^{-E({{\pvecsm}})\,(T-t)}\right)
  \nonumber\\ 
  & & C_{PA}(t,\pvec) =
  \sum\limits_{\xvecsm}e^{i{\pvecsm\cdot\xvecsm}} \langle
  \,P(t,\xvec)\, (A^I_0)^\dagger(0,\mbf{0})\,\rangle
  \simeq \frac{\mathcal{Z}_P\mathcal{Z}_A^\ast}{2E(\pvec)}
      \left(e^{-E({{\pvecsm}})\,t} -e^{-E({{\pvecsm}})\,(T-t)} \right)
  \,, \label{eq:twopoint}
\eea
where $T$ denotes the temporal extent of the lattice and where we have
already indicated the asymptotic behaviour for large Euclidean time
separations with the ground state matrix
elements 
\be
  \mathcal{Z}_{P}=  \left\langle 0 \left| P(0)
  \right|\pi(\pvec)\right\rangle\,, \quad
  \mathcal{Z}_{A}=  \left\langle 0 \left| (A_0^I)(0)
  \right|\pi(\pvec) \right\rangle.
\label{eq:zfactor}
\ee
The electromagnetic form factor is extracted from the
three-point function
\bea
& &  C_3(t,t_f,\pvec_i,\pvec_f) =
  \sum\limits_{\xvecsm_f,\xvecsm}
  e^{i{\pvecsm}_f\cdot({\xvecsm}_f-{\xvecsm})} 
  e^{i{\pvecsm}_i\cdot{\xvecsm}} \: \langle\,
  P(t_f,\xvec_f)\, V_0^I(t,\xvec)\,P^\dagger(0,\mbf{0})\,\rangle
  \nonumber \\
 & & \hspace{3mm} \simeq
  \frac{|\mathcal{Z}_P|^2}{4\:E(\pvec_i)\:E(\pvec_f)} \:
  \left\langle \pi(\pvec_f) \vert V_0^I(0) \vert \pi(\pvec_i)
  \right\rangle   \label{eq:threepoint} \\ 
 & & \hspace{3mm} \times \left[ \Theta(t_f-t) \:
  e^{-E({\pvecsm}_i)\:t-E(\pvecsm_f)\:(t_f-t)} 
  - \Theta(t-t_f) \: e^{-E(\pvecsm_i)\:(T-t)-E(\pvecsm_f)\:(t-t_f)}
  \right] \;. \nonumber 
\eea
In these formulae the initial pion source is located 
{on the first timeslice},
while $t$ is the temporal position of the insertion of the vector
current {and $t_f$ corresponds to the } position of the 
sink.

The calculation of matrix elements of the $\rmO(a)$ improved vector
current in the presence of twisted boundary conditions merits special
attention, due to the presence of the derivative of the tensor current
in eq.\,\refeq{eq:Vimpr2}. The contribution from the term proportional
to $\cv$ to the three-point correlation function in
eq.\,\refeq{eq:threepoint} reads
\be
   C_{\rm PTP}(t,t_f,\pvec_i,\pvec_f) = a\sum_{\nu=0}^3
   \tilde\partial_\nu 
   \sum\limits_{\xvecsm_f,\xvecsm}
   e^{i\pvecsm_f\cdot\xvecsm_f}
   e^{-i(\pvecsm_f-\pvecsm_i)\cdot\xvecsm} 
   \:\langle\,P(t_f,\xvec_f)\,T_{0\nu}(t,\xvec)\,P^\dagger(0,\mbf 0)\,\rangle\,,
\ee
where $x_f=(t_f,\xvec_f)$ and $x=(t,\xvec)$. Since $T_{00}\equiv0$
only spatial derivatives yield non-vanishing contributions. Moreover,
when periodic boundary conditions are imposed in the spatial
directions, the discretised derivatives vanish exactly, owing to
translational invariance.

In the presence of twisted boundary conditions this is no longer
true. The appearance of phase factors implies that uncancelled
contributions from the spatial boundary arise. A few lines of algebra
then yield the expression for $C_{\rm PTP}$, i.e.
\bea
 & & C_{\rm PTP}(t,t_f,\pvec_i,\pvec_f) = \frac{1}{2}
     \sum\limits_{\xvecsm_f}
     e^{i\pvecsm_f\cdot\xvecsm_f}
     \sum_{m=1}^3
     \sum\limits_{\xvecsm}
     e^{-i(\pvecsm_f-\pvecsm_i)\cdot\xvecsm} 
     \:\langle\,P(t_f,\xvec_f)\,T_{0\,m}(t,\xvec)\,P^\dagger(0,\mbf 0)\,\rangle \nonumber\\ 
 & & \hspace{3mm}\times\left\{
     \left(e^{-i(\thvecsm_f-\thvecsm_i)_m}-1\right) \delta_{x_m,L-1} -
     \left(e^{i(\thvecsm_f-\thvecsm_i)_m}-1\right)  \delta_{x_m,0}
     \right\}\,,
\label{eq:tensorimpr}
\eea
where $x_m$ denotes the $m^{\rm th}$ component of $\xvec$, while
$\mbf\theta_i,\,\mbf\theta_f$ are the twist angles applied to the initial and
final pions, respectively.

Our results presented in the following section show that the
contribution of the improvement term ${\cv}C_{\rm PTP}$ to $C_3$ is
numerically very small. Moreover, by construction this term vanishes
for $Q^2=0$, where the form factor is constrained to $\fpipizero=1$ by
symmetry. Given these observations and the fact that no
non-perturbative determination of $\cv$ is available for the
two-flavour theory we decided to drop the vector current improvement
altogether.

\subsection{The pion form factor, the decay constant and the light 
quark mass}

From now on we set $t_f=T/2$ and drop the corresponding argument in
correlation functions.

Following ref.~\cite{Boyle:2007wg} one can extract $\fpipi$ from
ratios of correlation functions such as
\bea
 R_1(t,\pvec_i,\pvec_f) & = & 4\,\zv^{\rm eff}\, \sqrt{E(\pvec_i)\:E(\pvec_f)}
 \sqrt{\frac{C_{3}(t,\pvec_i,\pvec_f)\:C_{3}(t,\pvec_f,\pvec_i)}
            {C_{PP}(T/2,\pvec_i)\:C_{PP}(T/2,\pvec_f)}}\,, \nonumber \\
 R_2(t,\pvec_i,\pvec_f) & = & 2\,\sqrt{E(\pvec_i)\:E(\pvec_f)}
 \sqrt{\frac{C_{3}(t,\pvec_i,\pvec_f)\:C_{3}(t,\pvec_f,\pvec_i)}
            {C_{3}(t,\pvec_i,\pvec_i)\:C_{3}(t,\pvec_f,\pvec_f)}}.
 \label{eq:ratios}
\eea
While any multiplicative renormalisation of the vector current cancels
in $R_2$, the ratio $R_1$ is renormalised by the factor $\zv^{\rm
eff}$. The form factor $\fpipi$ can then be determined via
\be
\label{eq:asympt}
R_k(t,\pvec_i,\pvec_f) = \fpipi \:
\left(E(\pvec_i)+E(\pvec_f)\right)\,,\quad k=1, 2,
\ee
where it should be kept in mind that this relation is valid for the
time component of the vector current and up to corrections induced by
excited state contributions. The renormalisation factor $\zv^{\rm
eff}$ of the vector current has been determined by imposing electric
charge conservation, which implies $\fpipizero=1$, at every value of
the lattice spacing. At the non-perturbative level $\zv^{\rm eff}$ is
obtained by evaluating
\be
\label{eq:zvextr}
\zv^{\rm eff} = \frac{C_{PP}(T/2,\mbf{0})}{2\:C_3(t,\mbf{0},\mbf{0})} \;.
\ee

We have also computed the renormalised and $\rmO(a)$ improved current
quark mass defined by the PCAC relation
\be
\label{eq:PCAC}
  \hat{m} \equiv \frac{\za}{\zp} 
  \left(1+\left[\ba-\bp\right]am_q\right)  m_{\rm PCAC},
  \quad m_{\rm PCAC} = \frac{1}{2}  \frac{\langle{\tilde
  \partial_0A^I_0(x)P^\dagger(0)}\rangle}{\langle{P(x)P^\dagger(0)}\rangle}
  \,,  
\ee
where the bare subtracted quark mass is given by
\be
  \label{eq:baremass}
  am_q = \frac{1}{2}  \left(\frac{1}{\kappa} - \frac{1}{\kappa_c}
  \right) \;. 
\ee
For the critical hopping parameter $\kappa_c$ we use the results
listed in table~\ref{tab:ensembleproperties}. The difference
$(\ba-\bp)$ is taken from~\cite{Fritzsch:2010aw}, and for $\za$ and
$\zp$ we use the non-perturbative results from \cite{Fritzsch:2012wq},
which update the earlier determinations from \cite{DellaMorte:2008xb}
and \cite{DellaMorte:2005kg}. The numerical values of the improvement
coefficients and renormalisation factors $\za$ and $\zp$ 
are listed table~\ref{tab:coeffs}.

\begin{table}
\begin{center}
\begin{tabular}{ccccc}
\hline\hline\\[-3mm]
$\beta$	& $\za$~\cite{Fritzsch:2012wq,DellaMorte:2008xb} &
$\zp$~\cite{Fritzsch:2012wq,DellaMorte:2005kg} &
$\ca$~\cite{DellaMorte:2005se} & $\ba-\bp$~\cite{Fritzsch:2010aw} \\
\hline
5.2&0.771(6)&0.518(5)& $-0.0641$ & $-0.1079$ \\
5.3&0.778(9)&0.518(5)& $-0.0506$ & $-0.0992$ \\
5.5&0.793(5)&0.518(5)& $-0.0361$ & $-0.0848$ \\
\hline\hline
\end{tabular}
\end{center}
\caption{\small Non-perturbative estimates of renormalisation factors
  and improvement coefficients, as used in our analysis.}
\label{tab:coeffs} 
\end{table}

We use the definition 
\be
  \label{eq:fpirat}
  \Fpi^{\rm bare} = \sqrt{2\,\left|\mathcal{Z}_{\rm P}\right|^2} \:
  \frac{m_{\rm PCAC}}{m_\pi^2} \;.
\ee
for the pion decay constant (e.g.~\cite{DelDebbio:2007pz}).
Note, that our normalisation for pseudoscalar decay constants
corresponds to a physical value of
$\Fpi=92.2\,\MeV$~\cite{Nakamura:2010zzi}.

\subsection{Ratios of correlation functions and excited states} 
\label{sec:excited}
\label{sec:exst}

As explained in the previous sections the pion form factor can be determined 
from the asymptotic form of the ratios~(\ref{eq:ratios})
of two- and three-point correlation functions (\ref{eq:twopoint}) 
and~(\ref{eq:threepoint}), i.e. for
large Euclidean time separations between the operator insertions. 

One can study the contribution of excited states to the ratios analytically by
inserting the spectral decomposition of the two- and three-point functions. 
We have looked at terms up to and including the first excited state which 
causes exponentially suppressed 
deviations from the constant behaviour expected for the ground state 
(cf. also the study in \cite{Brommel:2006ww}). 
The behaviour is the same for $R_1$ and $R_2$.
We control these contributions by 
choosing $t_f$ sufficiently large for all our measurements and
by tuning the fit window for every individual result for the ratio such that
exponential contaminations are sufficiently decayed.

There is a further, time-independent contamination proportional to 
$e^{-\Delta t_f/2}$, where $\Delta$ is the energy gap between the ground- 
and first excited state. We are not able to remove this contribution in our 
analysis because we do not have data for different choices of $t_f$.
Under the assumption that $\Delta\approx 2m_\pi$ and for our simulation
parameters as summarised table~\ref{tab:ensembleproperties}, this contribution
is however highly suppressed and it is therefore safe to neglect it.

Note also that due to current conservation at $Q^2=0$, which implies
$\fpipizero=1$, the contribution from the first excited state cancels
exactly between numerator and denominator. For $Q^2>0$ the
cancellation is no longer exact, but since it is smoothly connected to
vanishing momentum transfer, it is reasonable to assume that excited
state contaminations are rather small in the region that is particularly
relevant for the determination of the charge radius.

We also note that techniques for a systematic reduction of excited state
contaminations have been developed and applied
in~\cite{Bulava:2011yz,Capitani:2012gj}. They will allow for a more
precise estimation of residual effects in future calculations of meson
form factors.

\section{Lattice simulation and results}\label{sec:Latticesetup}

\subsection{Simulation parameters}\label{sec:basic}

\begin{table}[t]
\begin{center}
\begin{tabular}{lcccccccc}
\hline\hline&&\\[-4mm]
&$N_{\rm cfg}$& $T\times L^3$&$\beta$&$r_0/a$&$\kappa_{\rm crit.}$
&$\kappa_{\rm sea}$&$m_\pi$[MeV]&$m_\pi L$\\
		\hline&&\\[-4mm]
A3    &132&$64\times 32^3$&5.2&6.15(6) &0.136055(4)&0.13580&470&6.0\\
A4    &175&               &   &        &           &0.13590&365&4.7\\
A5    &108&               &   &        &           &0.13594&310&4.0\\
\hline                                   
E4    &81 &$64\times 32^3$&5.3&7.26(7) &0.136457(4)&0.13610&605&6.2\\
E5    &119&               &   &        &           &0.13625&450&4.6\\
F6    &233&$96\times 48^3$&   &        &           &0.13635&325&5.0\\
F7    &250&               &   &        &           &0.13638&280&4.3\\
\hline                                   
N3    &98 &$96\times 48^3$&5.5&10.00(11)&0.1367749(8)&0.13640&630&7.6\\
N4    &117&               &   &         &            &0.13650&535&6.5\\
N5    &189&               &   &         &            &0.13660&425&5.2\\
		\hline\hline&&
\end{tabular}
\caption{\small Summary of properties of the gauge ensembles. The results for
  the Sommer scale $r_0/a$ and for the critical hopping parameter
  $\kappa_{\rm{crit.}}$ are taken from \cite{Fritzsch:2012wq}. We also
  list approximate values for the pion masses in physical units.} 
\label{tab:ensembleproperties}
\end{center}
\end{table}

All our calculations are based on the CLS~\footnote{{\tt
https://twiki.cern.ch/twiki/bin/view/CLS/WebHome}} ensembles generated
with two dynamical flavours of non-perturbatively $\rmO(a)$ improved
Wilson fermions. The simulation parameters, i.e. the bare coupling and
the hopping parameter are listed alongside the values of some basic
observables in table~\ref{tab:ensembleproperties}. Gauge ensembles
were generated using the
DD-HMC~\cite{Luscher:2005rx,Luscher:2007se,Luscher:2007es} and
MP-HMC~\cite{Marinkovic:2010eg} algorithms.

To convert to physical units we use the Sommer scale
$r_0$~\cite{Sommer:1993ce}, which was recently determined on the CLS
ensembles~\cite{Leder:2010kz,Fritzsch:2012wq}. By computing the kaon
decay constant in units of $r_0$, i.e. $(f_{\rm{K}}\,r_0)$, taking the
continuum limit and combining it with the experimental value of
$f_{\rm{K}}$, one obtains
$r_0=0.503(10)\,\fm$~\cite{Fritzsch:2012wq}. This value is consistent
with the scale setting procedure based on the mass of the
$\Omega$-baryon described in~\cite{Capitani:2011fg}, provided that the
updated results for $r_0$ from~\cite{Fritzsch:2012wq} are used. Hence, the
three lattice spacings of the ensembles used here are in the range 
0.05fm -- 0.08fm.

We have evaluated all two- and three-point correlation functions
using two hits of stochastic $Z_2\times Z_2$ wall
sources~\cite{Foster:1998vw,McNeile:2006bz,Boyle:2008rh,Endress:2011jc}.
For each configuration subsequent hits were evaluated on two different
timeslices which we separated in time by $T/2$, except for ensemble A5
where we applied hits on four timeslices separated by $T/4$. The pion
mass, the pion decay constant, the PCAC quark mass and the pion vector
form factor were evaluated in this way on the ensemble of gauge
configurations as summarised in table~\ref{tab:basicmeasurements}.
Our ensembles cover a large range of quark masses and lattice
spacings. As a safeguard against large finite-size effects, we have
kept $m_\pi L\geq4$ on all ensembles. Our parameter choice thus allows
for a comprehensive study of systematic effects, relating to chiral
and continuum extrapolations.

As motivated above, we are particularly interested in the region of
small momentum transfers. We have therefore tuned the twist angles
specifically to achieve a high resolution for small values of $Q^2$,
using eq.\,\refeq{eq:qsq} and a first rough determination of the pion
mass on all ensembles. Twisted boundary conditions were applied in the
$x$-direction to the pions in both the initial and final states,
whilst projecting on vanishing Fourier momenta,
$\pvec_i=\pvec_f=0$. Table~\ref{tab:twistangles} contains the full set
of angles used in our simulations. Correlation functions for the pion
vector form factor were generated simultaneously with those required
for the computation of the $K\to\pi$ semi-leptonic form factor. For
the latter, partially twisted boundary conditions allow for simulating
directly at the phenomenologically relevant kinematical point of
vanishing momentum transfer~\cite{Boyle:2007wg,Boyle:2010bh} between
the kaon and the pion. The angles $\theta_1,\ldots,\theta_4$ in
table~\ref{tab:twistangles} have been tuned such as to realise $Q^2=0$
for the $K\to\pi$ matrix elements. We added one extra twist angle
$\theta_5$ which was chosen such as to yield a dense set of data
points for small momentum transfers.

\begin{table}
\begin{center}
\begin{tabular}{lcccccc}
\hline\hline&&\\[-4mm]
set&$\theta_0$& $\theta_1$& $\theta_2$& $\theta_3$& $\theta_4$& $\theta_5$\\
\hline&&\\[-4mm]
A4&$0.0$&$\pm2.2658$&$\pm1.8438$&$\pm1.3582$&$\pm0.7745$&$\pm2.5$\\[0mm]
A5&$0.0$&$\pm2.4380$&$\pm2.0281$&$\pm1.5544$&$\pm0.9777$&$\pm2.5$\\[0mm]
E4&$0.0$&$\pm1.6799$&$\pm1.2748$&$\pm0.8195$&$\pm0.2969$&$\pm2.5$\\[0mm]
E5&$0.0$&$\pm2.1728$&$\pm1.7866$&$\pm1.3476$&$\pm0.8344$&$\pm2.5$\\[0mm]
F6&$0.0$&$\pm3.2455$&$\pm2.9371$&$\pm2.6028$&$\pm2.2355$&$\pm1.5$\\[0mm]
F7&$0.0$&$\pm3.7892$&$\pm3.5196$&$\pm3.2323$&$\pm2.9231$&$\pm2.0$\\[0mm]
N3&$0.0$&$\pm0.7538$&$\pm0.3777$&$\pm0.4935$&$\pm1.0146$&$\pm4.0$\\[0mm]
N4&$0.0$&$\pm1.2730$&$\pm0.8936$&$\pm0.4726$&$\pm0.5443$&$\pm3.9$\\[0mm]
N5&$0.0$&$\pm1.7513$&$\pm1.3942$&$\pm0.9945$&$\pm0.5311$&$\pm3.2$\\[0mm]
\hline
\hline&&
\end{tabular}
\caption{\small The choice of twist angles applied to the initial and final
  mesons in the $x$-direction.}
\label{tab:twistangles}
\end{center}
\end{table}

On each of the ensembles listed in table~\ref{tab:ensembleproperties}
we have computed the ratios $R_1$ and $R_2$ in eq.\,\refeq{eq:ratios}
for all possible combinations of twists applied to the valence quark
being probed by the vector current.

\begin{figure}
\begin{center}
 \begin{center}
  \includegraphics[width=0.4\textwidth]{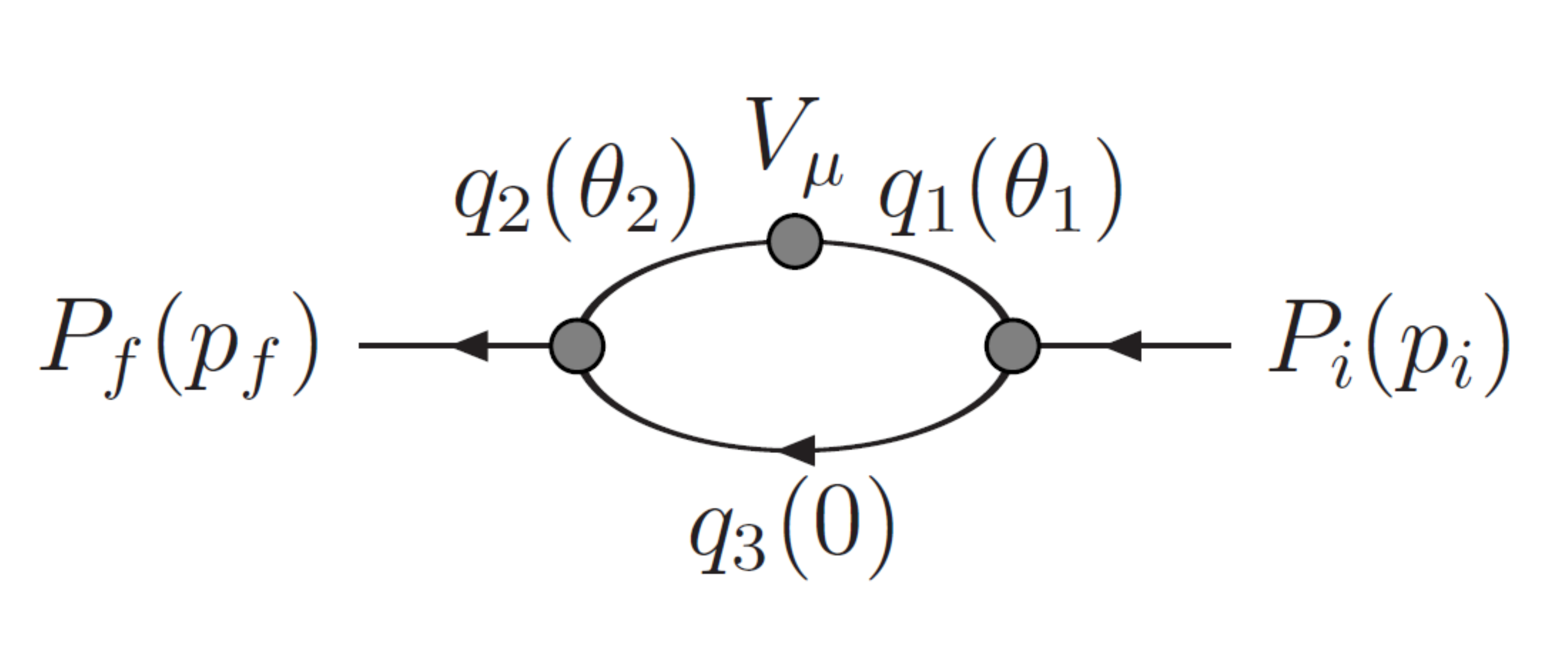}
  \end{center}
\caption{\small Graphical representation of 3pt-function $C_3$ with
  explanation of twist angles.}
\label{fig:quarkflow}
\end{center}
\end{figure}

\subsection{Data analysis and fitting procedure}\label{sec:fitting}

All error estimates are computed by resampling using 
the bootstrap procedure~\cite{efron:1979} with 1000 bins. 
Masses, decay constants, form factor and other quantities have been
extracted from correlation functions via suitable fits. While
simulation data from a given ensemble are correlated, it is
often difficult to obtain sufficiently precise estimates of the full
covariance matrix based on a finite set of gauge configurations. As a
consequence, numerical instabilities can occur in the least-square
minimisations (see for example~\cite{Michael:1993yj}). We have
therefore chosen to quote our main results from uncorrelated fits. 

The quantities which we fitted to ChPT often contain input parameters such
as renormalisation factors, which have their own intrinsic uncertainties. In
order to take the latter into account we have folded them into our analysis
via the following procedure: first we generated a pseudo-bootstrap
distribution with 1000 bins for the input quantity, whose width was designed
such that it reproduced the quoted uncertainty. We checked explicitly that the
bootstrap error obtained for the combination of distributions was compatible
with the corresponding estimate determined via the usual error propagation.

This procedure was also applied in combinations such as $m_\pi r_0$, despite
the fact that the determination of $r_0$ was mostly performed on the same
ensembles. However, since the set of configurations used to compute the pion
form factor did not exactly coincide with those used in the calculation of
$r_0$, we chose to ignore the partial correlation of $r_0$ with our data,
although this results in a larger overall uncertainty.

\subsubsection{Pion mass and decay constant}

We extracted the pion energy from a cosh-fit to the two-point
function $C_{\rm PP}(t,\pvec)$, after checking that the results with our
choice of fit-ranges remain unchanged when a three-pion state is included as
the first excited state, as suggested in~\cite{DelDebbio:2007pz}. We obtained
the current quark mass $m_{\rm PCAC}$ from a constant fit to 
the ratio in eq.~\refeq{eq:PCAC}. The bare pion decay constant was then
determined from eq.\,\refeq{eq:fpirat}, using the result for 
$\mathcal{Z}_P$ and $m_\pi$ from the above cosh-fit and $m_{\rm PCAC}$.
%
\begin{table}
\begin{center}
\begin{tabular}{lccccc}
\hline\hline&&\\[-4mm]
ensemble & $r_0\:m_\pi$ & $r_0\:\fpi$ & $r_0\:\hat{m}$ & $\zv^{\rm{eff}}$ &
$\rsq/r_0^2$ \\
\hline\hline&&\\[-4mm]
A3 & 1.161(12) & 0.280( 8) & 0.090(3) & 0.73228( 7) & 1.14 ( 5) \\
A4 & 0.895(11) & 0.247(12) & 0.052(3) & 0.72885(12) & 1.20 ( 7) \\
A5 & 0.761(11) & 0.251(14) & 0.040(2) & 0.72731(10) & 1.48 ( 9) \\
\hline                                                    
E4 & 1.406(16) & 0.287(10) & 0.128(5) & 0.74962( 8) & 0.98 ( 4) \\
E5 & 1.048(13) & 0.271(11) & 0.078(4) & 0.74461( 8) & 1.18 ( 5) \\
F6 & 0.752( 8) & 0.254( 8) & 0.041(2) & 0.74119( 4) & 1.37 ( 6) \\
F7 & 0.646( 7) & 0.237( 8) & 0.029(1) & 0.74030( 5) & 1.61 (10) \\
\hline
N3 & 1.593(18) & 0.329( 7) & 0.188(5) & 0.77162( 3) & 0.90 ( 3) \\
N4 & 1.360(16) & 0.304( 9) & 0.139(4) & 0.76855( 3) & 1.04 ( 3) \\
N5 & 1.080(13) & 0.291( 8) & 0.091(3) & 0.76543( 3) & 1.17 ( 4) \\
\hline\hline
\end{tabular}
\caption{\small Results for basic quantities. }  
\label{tab:basicmeasurements}
\end{center}
\end{table}
%
The fit-results  in units of $r_0$ are summarised in
table~\ref{tab:basicmeasurements}. They are in agreement with the
results obtained in~\cite{Fritzsch:2012wq}\footnote{Note that the analysis 
in~\cite{Fritzsch:2012wq} was done using different numbers of measurements,
source positions and also fitting strategies.}.

\begin{figure}[ht]
\begin{center}
 \includegraphics[width=0.9\textwidth]{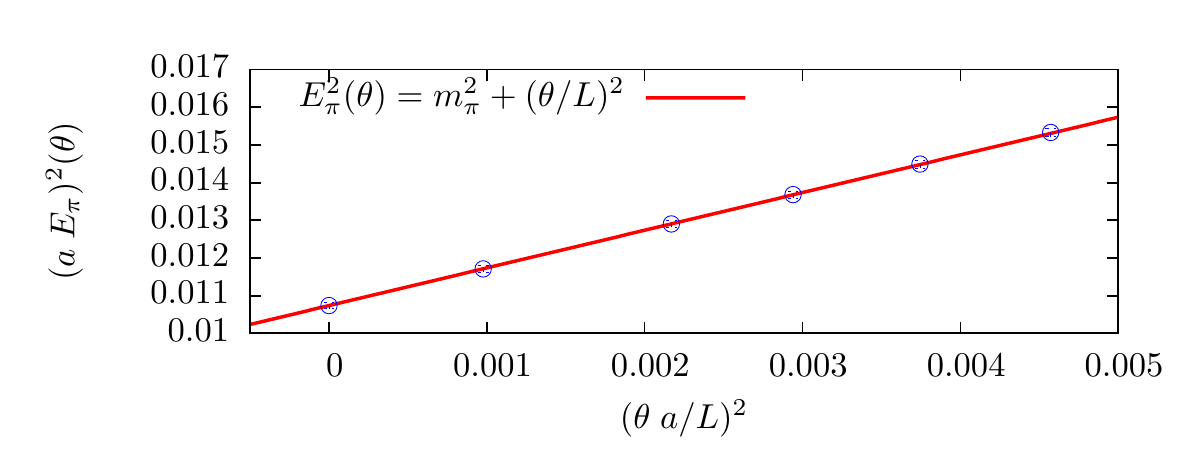}
\vspace{-0.5cm}
\caption{\small The pion dispersion relation with partially twisted boundary
conditions compared to the continuum dispersion relation, eq.~\refeq{eq:dispersion}. {The vertical axis shows the interval of squared pion momenta
up to about 40\,MeV$^2$.}}
\label{fig:dispersion}
\end{center}
\end{figure}

Up to cut-off and exponentially suppressed finite-volume effects, 
the pion energy obeys the dispersion relation~(\ref{eq:dispersion})
where $\pvec$ is the difference of the twist angles applied to the
pion's valence
quarks~\cite{Sachrajda:2004mi,Flynn:2005in,Jiang:2006gna} divided by
the spatial extent $L$ of the lattice. As an example for how well the
continuum dispersion relation is reproduced by our data we show the
numerical results on ensemble F6 together with eq.~(\ref{eq:dispersion})
in figure~\ref{fig:dispersion}. Note that for these small momenta the
difference between continuum and lattice dispersion relations is
negligible. In the remainder of the
analysis we use eq.~(\ref{eq:dispersion}),
i.e. we always determine the kinematics in terms of the pion energy at
rest together with the exactly know twist angles.
\subsubsection{Form factor}\label{sec:form factor}
As discussed in detail in earlier sections we determine the form factor
from the plateau of suitable ratios of two- and three-point functions, 
$R_1$ and $R_2$, as defined in eq.~(\ref{eq:ratios}). 
For each ensemble around 60 kinematical points 
entered our analysis. We inspected every single plateau by eye for both
ratios and then determined the fit range. Some of the plateaus, in particular
for larger induced momentum were found to be of unsatisfactory quality. We 
discarded those ratios from the subsequent analysis. 
Generally the data for $R_1$ was of better quality and we decided to 
use only results from this ratio in the following. We confirmed however,
that using  $R_2$ instead leads to the same results and conclusions.
\begin{figure}[t]
 \centering
\includegraphics[width=.9\textwidth]{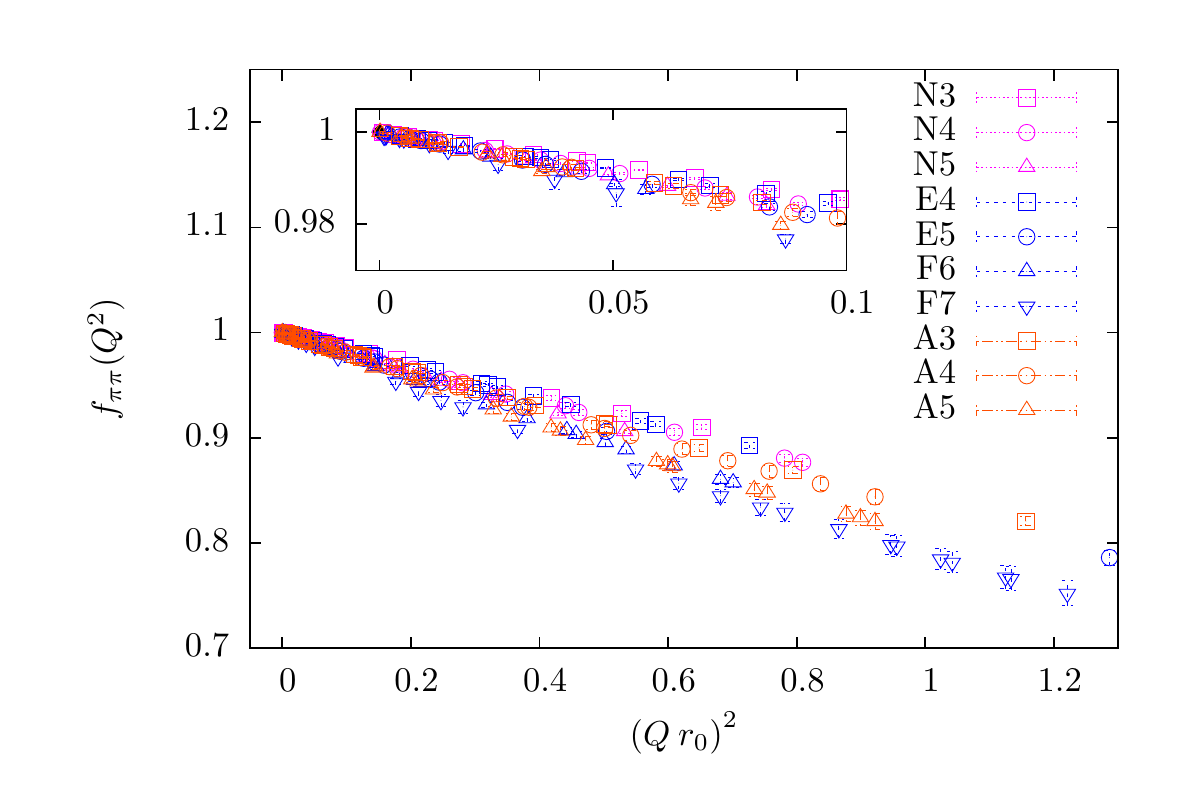}
\vspace{-0.5cm}
 \caption{\small Results for the pion form factor for all ensembles. The inset
  shows a zoom into the region of small $Q^2$.}
 \label{fig:fpipi_overview}
\end{figure}
		
Figure~\ref{fig:fpipi_overview} shows our results for the form factor 
for all ensembles. The small inset shows the region of very
small $Q^2$ which we are concentrating on in this work. 
\begin{figure}[t]
 \centering
\includegraphics[width=.9\textwidth]{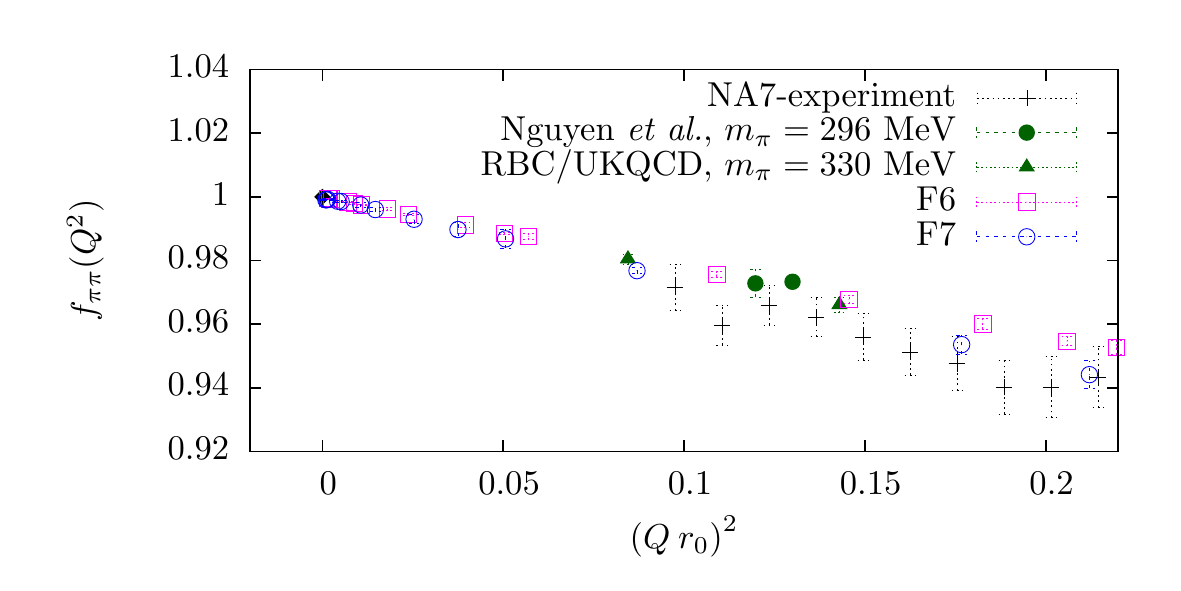}
\vspace{-0.5cm}
 \caption{\small Compilation of results for the pion form factor in dynamical
  lattice QCD~\cite{Boyle:2007wg,Boyle:2008yd,Frezzotti:2008dr,Nguyen:2011ek}
  and as determined from experiment~\cite{Amendolia:1986wj}.}
 \label{fig:fpipi_comparison}
\end{figure}
In figure~\ref{fig:fpipi_comparison} we compare our results for the
two lightest pions (ensembles F6 and F7) to the experimental
measurement, as well as to results of two other lattice collaborations
in the region of small $Q^2$. The plot illustrates
nicely that partially twisted boundary conditions are a powerful method for
isolating the low-momentum behaviour of the form factor.
\subsection{Finite volume corrections}\label{sec:fvol}
Finite volume effects for the quantities
considered here are expected to be suppressed exponentially $\propto e^{-m_\pi L}$. In order to keep these effects small we
have used only ensembles for which $m_\pi\,L\geq4$ (cf. 
table~\ref{tab:ensembleproperties}) but we have also made the effort
to remove residual effects systematically.
This was not possible directly
using simulation data since we do not have results for different volumes at 
fixed pion mass at our disposal. 
Instead we have used predictions of chiral perturbation theory.
The corresponding expressions have been derived
in~\cite{Colangelo:2005gd} for the pion mass and the pion decay
constant. Remarkably, the full expressions at NNLO contain only the
low-energy constants (LECs) which appear at NLO, which has been referred to
as the ``elevator-effect'' in~\cite{Colangelo:2005gd}. For the vector
form factor with partially twisted boundary conditions, finite-volume
effects have been computed in NLO chiral perturbation
theory~\cite{Jiang:2006gna}, for the case that either $\mbf\theta_i=0$
or $\mbf\theta_f=0$, and in~\cite{Jiang:2008te} for the
Breit-frame~$\mbf\theta_i=-\mbf\theta_f$.

In order to evaluate finite-volume effects in ChPT, we initially fixed the
relevant LECs in the same way as Colangelo, D\"urr and
H\"afeli~\cite{Colangelo:2005gd}, who took their values
from~\cite{Colangelo:2001df}. To become independent of external input
quantities, we performed the following iterative procedure: first, we
applied the finite-volume correction based on ChPT and the LECs
from~\cite{Colangelo:2001df} as input. The subsequent fits
of the chiral behaviour of the lattice data to the expressions of ChPT
described in the following section provide us with predictions for the
LECs from which the finite-volume shift is
re-computed. After repeating this procedure twice no significant
change in the output LECs was observed. We note that
the estimates after the final iteration are
well compatible with the values of~\cite{Colangelo:2001df}.

Despite the fact that the ChPT estimates of finite-volume effects turn
out to be numerically small, we apply these corrections to the lattice
data prior to any subsequent analysis. In the case of the form factor
we restrict ourselves in the following to only those kinematical situations 
where predictions of finite volume effects are available.
\subsection{Extraction of the charge radius}\label{sec:chrad}
%
\begin{figure}
 \centering
\includegraphics[width=.9\textwidth]{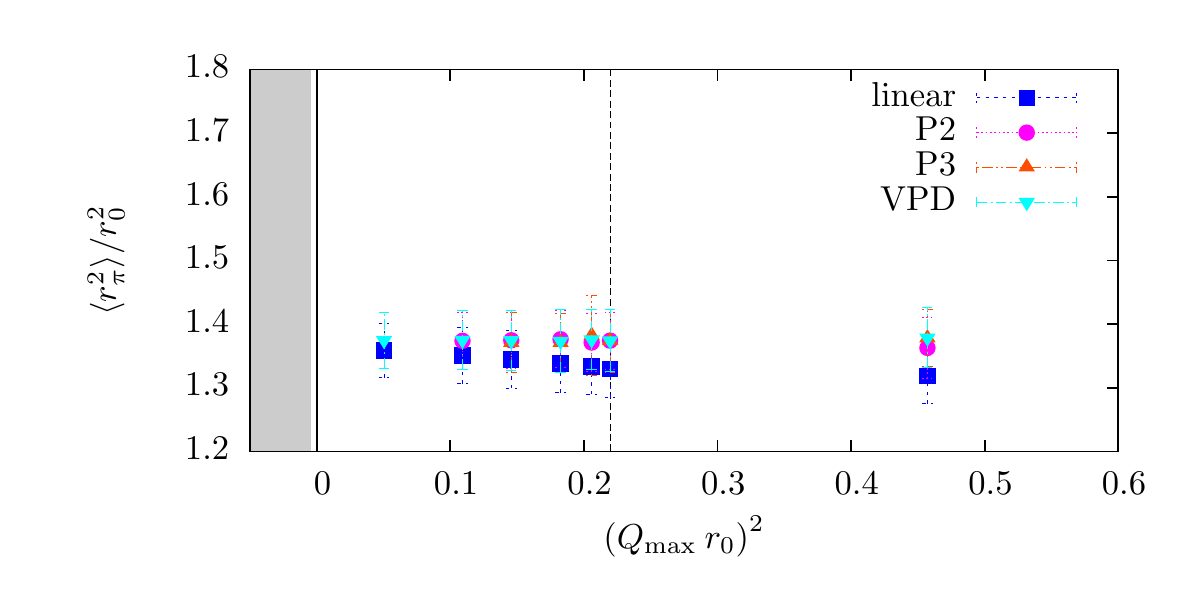}
\vspace{-0.5cm}
 \caption{\small The squared charge radius for a pion mass of about
 $325\,\MeV$ (ensemble~F6), plotted versus the maximum value of the
 $Q^2$-interval entering the fit, in units of $r_0$. Labels P2 and P3
 denote the results of a fit to polynomials of degree two and three,
 respectively, and the results denoted by VPD are the results of a fit
 to the form in eq.~\refeq{eq:vpd-model}.}
 \label{fig:rsq_overview}
\end{figure}
%
The charge radius of the pion is defined as the derivative of the form
factor with respect to the momentum transfer at $Q^2=0$
(c.f. eq.\,\refeq{eq:chrad}). In practice, and this also affects the determination
from experimental data for the form factor, one fits a model for the 
$Q^2$--dependence to the data (e.g. pole- or polynomial ansatz) and extracts
the charge radius in terms of the slope at the origin.
Until recently, data from both experiment and lattice~QCD did not cover the
region of very low momentum transfer $Q^2<0.013$~GeV$^2$, which is where one
would ideally like to extract the slope. Studies of the systematics
introduced by the fit-{\it ansatz} were therefore very limited.

The high density of data points for $\fpipi$ near $Q^2=0$ -- shown for
all ensembles in figure~\ref{fig:fpipi_overview} -- allows us to
constrain the functional form of the form factor very accurately and
to reduce any model dependence in the extraction of the charge radius
to a minimum. In practice we compare radii as extracted from linear
fits in $Q^2$, polynomial fits, as well as pole fits of the form
\be
\label{eq:vpd-model}
\left. \fpipi \right|_{\text{VPD}} = \left( 1-\frac{\displaystyle
\rpi_\text{VPD}}{6}\:Q^2 \right)^{-1} \;.
\ee
The latter was already employed in~\cite{Amendolia:1986wj} to determine
the charge radius from experimental data and is usually used in
the determination from lattice data
(e.g.~\cite{Boyle:2007wg,Boyle:2008yd,Frezzotti:2008dr,Nguyen:2011ek}).

Figure~\ref{fig:rsq_overview} shows a representative example of the
charge radius on ensemble F6, which corresponds to a pion mass of
$325\,\MeV$. Data points were obtained by fitting the $Q^2$-dependence
of the form factor to a particular ansatz within an interval $0 \leq
Q^2\leq Q_{\rm{max}}^2$. The resulting estimates for the squared
charge radius are then plotted versus the value of $Q_{\rm max}^2$
used in the fit. In order to compare results for different mass and
lattice spacing, we express all dimensionful quantities in units
of~$r_0$. In the regime of low $Q^2$, one observes good agreement
between different types of fits. Interestingly, higher orders in $Q^2$
turn out to become relevant very early as can be seen from an
increasing discrepancy between the linear fit on the one hand, and the
polynomial and pole fits on the other.

\begin{figure}
 \centering
\includegraphics[width=.9\textwidth]{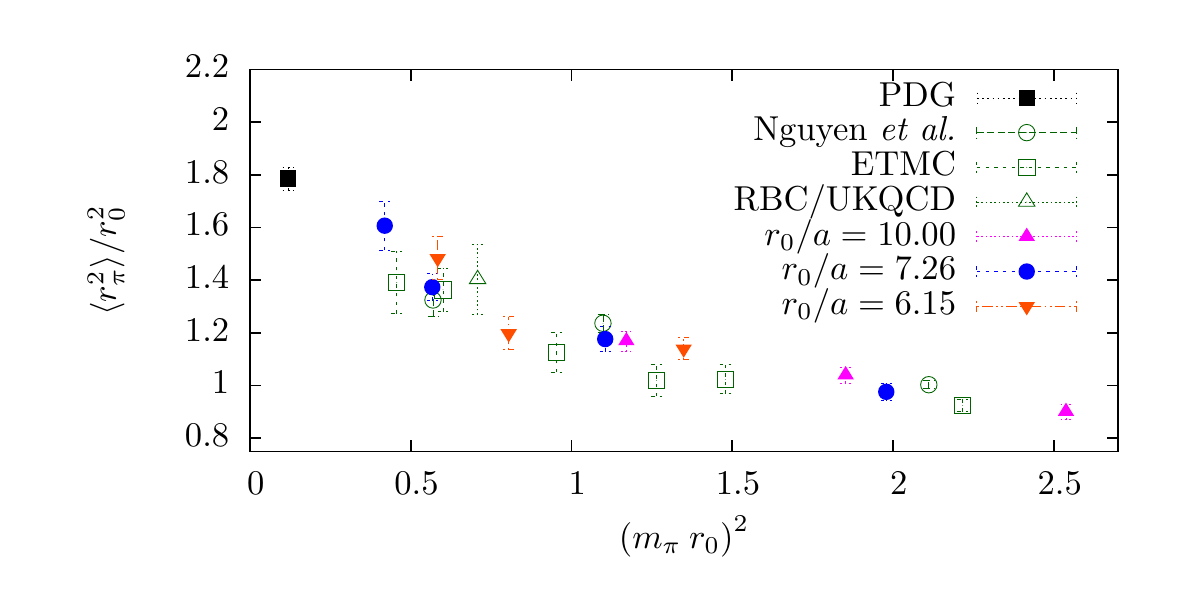}
\vspace{-0.5cm}
 \caption{\small Compilation of results for the pion charge radius in
   dynamical lattice
   QCD~\cite{Boyle:2007wg,Boyle:2008yd,Frezzotti:2008dr,Nguyen:2011ek}
   and the value quoted by the particle data group~\cite{Nakamura:2010zzi}.
   Most of the lattice data is extracted from a single pole fit except for the
   data from RBC/UKQCD and the data from this study, for which we show the
quadratic fit based on an identical $Q^2$-cut at
$(Q_{\rm{max}}\,r_0)^2\approx0.034$~GeV$^2$.}
 \label{fig:chrad_overview}
\end{figure}

In the following we will use the result of the fit using a second
order polynomial (P2), imposing a cut at
$(Q_{\rm{max}}\,r_0)^2\approx0.22$, which in physical units
corresponds to about $0.034$~GeV$^2$. In this range all our {\it
ans\"atze} are mutually compatible. We prefer the second-order
polynomial over the linear fit, since it yields consistent results
over a larger range of $Q^2_{\rm max}$. The results for $\rsq$ on all
ensembles are listed in table~\ref{tab:basicmeasurements}, and a
comparison with results from other collaborations is provided in
figure~\ref{fig:chrad_overview}.

Our lattice data suggest that the form factor can be represented very
well by a second-order polynomial up to values of the momentum
transfer which have been probed by the NA7-experiment~\cite{Amendolia:1986wj}. 

\section{Chiral and continuum extrapolations}\label{sec:extrapolations}
Table~\ref{tab:basicmeasurements} summarises our results for the pion mass,
the pion decay constant and the charge radius. We refrain from presenting the
abundant numerical data on the form factor itself but will use it for some of
the discussions that follow. In this section we present our attempts at
parameterising the lattice data and also at extrapolating it to the physical
point, i.e. to the physical quark mass and to the continuum and infinite
volume limits.
\subsection{Fits guided by chiral perturbation theory}
Chiral perturbation theory~\cite{Gasser:1983yg,Gasser:1984gg,Gasser:1984ux}
provides a comprehensive effective theory framework for describing the
low-energy dynamics of QCD. Its predictions for the functional form of the
mass-, momentum- and cutoff dependence of low-energy observables are a
standard tool in lattice QCD for extrapolating lattice data in parameter space
(volume, quark mass, lattice spacing, momentum).  For the two flavour theory
the expressions for the pion mass, the decay constant and the pion form
factor, and consequently also its charge radius, have been computed within
this framework at NLO~\cite{Gasser:1983yg,Gasser:1984ux} and at
NNLO~\cite{Burgi:1996qi,Burgi:1996mm,Bijnens:1997vq,Bijnens:1998fm}. We
summarise the corresponding formulae together with our parameterisation of
cutoff effects in appendix~\ref{app:chiptA} and \ref{app:chiptB}. At each
order in the expansion new mass- and momentum-independent LECs appear. They
are \textit{a priori} unknown parameters, unconstrained by symmetry, yet they
can be determined from lattice QCD data (for a summary of recent results see
ref.~\cite{Colangelo:2010et}). Some LECs contribute to the chiral expansion of
more than one quantity which can be exploited for correlations and consistency
checks. These correlations motivate simultaneous analyses of more than one
observable to gain better control over the chiral extrapolations. Here we
compare the following fits:\\[1mm]
\hspace{5mm}
\begin{tabular}{lll}
\textbullet \;\;individual fits&to $m_\pi^2$, $F_\pi$, $f_{\pi\pi}$ and
		$\rpi$& at NLO\\
\textbullet \;\;joint fits &to $(m_\pi^2, F_\pi)$ &at NLO and NNLO\\
\textbullet \;\;joint fits &to $(m_\pi^2, F_\pi, f_{\pi\pi})$  &at NLO
and NNLO\\ 
\textbullet \;\;joint fits &to $(m_\pi^2, F_\pi, \rpi)$ &at NLO and
NNLO. 
\end{tabular}\\[1mm]
Since the chiral series is expected to provide a good representation of QCD
only up to a certain low-energy scale we repeat all fits three times
including, respectively, all data points for pion masses up to about 430\,MeV,
560\,MeV and 630\,MeV, while monitoring whether the results depend on the
choice of the mass cutoff. We have to be less worried about the range of the
momentum transfers which enter the fits: The use of partially twisted boundary
conditions provides us with many data points well within the realm of chiral
perturbation theory. We have extracted the charge radius at very small
momentum transfers of up to $(r_0 Q)^2=0.22$ which in physical units
corresponds to about (190~MeV)$^2$. For fits to the form factor our choice
for the momentum cut is $(r_0 Q)^2=0.1$ (about (120~MeV)$^2$) for NLO fits,
where the $Q^2$-dependence of $f_{\pi\pi}$ is mostly linear (cf.
figure~\ref{fig:rsq_overview}),
and $(r_0 Q)^2=0.5$ (about (300~MeV)$^2$) for NNLO fits.

Before discussing the various fits we have performed, it is instructive to
recall earlier determinations of low-energy parameters in lattice QCD. The
FLAG review \cite{Colangelo:2010et} quotes global estimates of
$F_\pi/F=1.073(15)$ and $\lc{3}=3.2(8)$, while typical results for the LECs
$\lc{4}$ and $\lc{6}$ can be summarised as $\lc{4}\approx4$ and
$\lc{6}=12-16$, respectively.
\subsubsection{NLO chiral fits}
\begin{table}
\begin{center}
\footnotesize      
\begin{tabular}{llrrrrrrrrr}
\hline\hline &&&&&&&&&& \\[-3mm]
$m_\pi^{\rm cut}$&$\chi^2/dof$&$r_0F$&$r_0B$&$\lc{3}$&$\lc{4}$&$\lc{6}
$&$\alpha_m$&$\alpha_f$&$\alpha_r$\\
\hline\\[-3.5mm]
\bf NLO&\multicolumn{10}{l}{ $\bm F_\pi$}\\
all&0.8&0.225(15)&&&4.6(1)&&&$-$1.3(6)&\\
560\,MeV&0.6&0.226(16)&&&4.5(2)&&&$-$1.2(6)&\\
430\,MeV&0.5&0.220(17)&&&4.7(3)&&&$-$1.2(6)&\\
\bf NLO &\multicolumn{10}{l}{$\bm m_\pi^2$ fixed $F_\pi$}\\
all&2.1&&6.0(4)&2.6(5)&&&$-$5(1)&&\\
560\,MeV&2.0&&6.0(4)&2.3(6)&&&$-$5(1)&&\\
430\,MeV&1.2&&6.2(5)&3.0(8)&&&$-$4(1)&&\\
\bf NLO &\multicolumn{10}{l}{$\bm f_{\pi\pi}$}\\
all&1.0&0.135( 6)&&&&6.7(3)&&&$-$2(3)\\
560\,MeV&0.9&0.140( 7)&&&&7.0(4)&&&$-$2(3)\\
430\,MeV&0.8&0.142( 8)&&&&7.0(4)&&&$-$0(3)\\
\bf NLO &\multicolumn{10}{l}{$\bm \rpi$}\\
all&0.8&0.136( 6)&&&&6.7(3)&&&$-$1(3)\\
560\,MeV&0.9&0.137( 7)&&&&6.8(4)&&&$-$1(3)\\
430\,MeV&1.2&0.137(12)&&&&6.7(7)&&&0(3)\\
\bf NLO &\multicolumn{10}{l}{$\bm \rsq$ fixed $\bm F_\pi$}\\
all&6.8&&&&&12.0(5)&&&8(3)\\
560\,MeV&6.1&&&&&12.7(5)&&&4(3)\\
427\,MeV&4.1&&&&&13.6(7)&&&1(3)\\
\hline\hline
\end{tabular}
\caption{\small Fit results for individual observables based on ChPT at
  NLO. The coefficients $\alpha_m, \alpha_f$ and $\alpha_r$ parameterise
  lattice artefacts. For full expressions see
  appendix~\ref{app:chiptB}.}\label{tab:individualfits}
\end{center}
\end{table}
\begin{sidewaystable}
\footnotesize          \begin{tabular}{l@{\hspace{1mm}}l@{\hspace{1mm}}l@{\hspace{1mm}}l@{\hspace{1mm}}l@{\hspace{1mm}}l@{\hspace{1mm}}l@{\hspace{1mm}}l@{\hspace{1mm}}l@{\hspace{1mm}}l@{\hspace{1mm}}l@{\hspace{1mm}}l@{\hspace{1mm}}l@{\hspace{1mm}}l@{\hspace{1mm}}l@{\hspace{1mm}}l@{\hspace{1mm}}l@{\hspace{1mm}}l@{\hspace{1mm}}l@{\hspace{1mm}}l@{\hspace{1mm}}l@{\hspace{1mm}}l@{\hspace{1mm}}l@{\hspace{1mm}}l@{\hspace{1mm}}l}
\hline\hline\\[-3mm]
$m_\pi^{\rm cut}$&$\chi^2/dof$&$r_0F$&$r_0B$&$\lc{3}$&$\lc{4}$&$\lc{6}$
&$r_{V_1}\:10^4$&$r_{V_2 }\:10^5$&$r_f\:10^5$&$r_M\:10^5$&$\alpha_m$
&$\alpha_f$&$\alpha_r$\\
\hline\\[-3.5mm]
\bf NLO &\multicolumn{14}{l}{$\bm F_\pi$, $\bm m_\pi^2$}\\
all&1.4&0.221(15)&6.1(5)&2.7( 5)&4.7( 1)&&&&&&$-$4.7(11)&$-$0.9(5)&\\
560\,MeV&1.2&0.217(18)&6.0(5)&2.5( 6)&4.6( 2)&&&&&&$-$4.6(11)&$-$0.7(6)&\\
430\,MeV&0.8&0.213(17)&6.3(5)&3.0( 6)&4.7( 3)&&&&&&$-$4.3(12)&$-$0.7(6)&\\
\bf NNLO &\multicolumn{14}{l}{$\bm F_\pi$, $\bm m_\pi^2$}\\
all&1.6&0.228(14)&5.9(4)&2.3(12)&4.1( 6)&&&&6( 9)&$-$8(
8)&$-$4.5(11)&$-$0.8(5)&\\
560\,MeV&1.3&0.215(17)&6.1(5)&4.1(13)&5.3(
8)&&&&$-$17(14)&12(11)&$-$4.9(11)&$-$0.9(6)&\\
430\,MeV&1.1&0.221(33)&6.2(9)&3.3(72)&4.1(34)&&&&11(99)&1(126)&$-$4.3(12)&$-$0.7(6)&\\
\hline\\[-3.5mm]
\bf NLO &\multicolumn{14}{l}{$\bm F_\pi$, $\bm m_\pi^2$, $\bm f_{\pi\pi}$}\\
all&2.0&0.137( 7)&6.6(5)&3.1( 2)&4.6( 1)&6.9( 3)&&&&&$-$5.5(10)&1.3(3)&$-$5(3)\\
560\,MeV&2.0&0.145( 9)&6.5(5)&3.1( 2)&4.6( 1)&7.3(
4)&&&&&$-$5.3(10)&1.1(4)&$-$4(3)\\
430\,MeV&1.9&0.153(10)&6.5(5)&3.0( 3)&4.6( 1)&7.6( 5)&&&&&$-$5.1(10)&0.9(4)&$-$1(3)\\
\bf NLO &\multicolumn{14}{l}{$\bm F_\pi$, $\bm m_\pi^2$, $\bm \rsq$}\\
all&2.4&0.159(12)&6.4(5)&3.0( 3)&4.7( 1)&7.8( 6)&&&&&$-$5.0(10)&0.8(4)&0(3)\\
560\,MeV&2.0&0.150(15)&6.5(6)&2.9( 3)&4.6( 2)&7.5( 7)&&&&&$-$4.7(11)&1.1(5)&$-$3(4)\\
430\,MeV&1.6&0.180(25)&6.4(6)&3.0( 5)&4.8( 3)&9.4(16)&&&&&$-$4.3(12)&0.2(7)&$-$1(4)\\
\bf NNLO &\multicolumn{14}{l}{$\bm F_\pi$, $\bm m_\pi^2$, $\bm f_{\pi\pi}$}\\
all&1.2&0.235(18)&5.9(4)&2.2(12)&4.0(
6)&17.7(21)&$-$1.4(5)&8(4)&9(11)&$-$8(9)&$-$4.6(11)&$-$1.0(6)&$-$5(3)\\
560\,MeV&1.1&0.212(23)&6.1(5)&4.0(14)&5.4(
9)&15.3(23)&$-$0.9(4)&6(3)&$-$18(14)&11(14)& $-$4.8(11)&$-$0.8(6)&$-$5(3)\\
430\,MeV&0.7&0.147(29)&6.5(8)&4.0(21)&7.7(16)&10.2(16)&$-$0.09(25)&0.7(13)&$-$40(17)&4
(15)&$-$4.3(12)&$-$0.5(7)&$-$2(4)\\
\bf NNLO &\multicolumn{14}{l}{$\bm{F_\pi}$, $\bm m_\pi^2$, $\bm\rpi$}\\
all&1.4&0.226(11)&5.9(4)&2.3(12)&4.2(6)&16.8(12)&$-$1.2(3)&&6( 9)&$-$8(
8)&$-$4.5(11)&$-$0.8(5)&$-$5(3)\\
560\,MeV&1.3&0.213(14)&6.1(5)&4.0(13)&5.3(
8)&15.5(15)&$-$1.0(3)&&$-$18(13)&11(12)&$-$4.9(11)&$-$0.9(5)&$-$5(3)\\
430\,MeV&1.2&0.197(47)&6.3(
9)&3.5(55)&5.9(32)&13.5(38)&$-$0.5(7)&&$-$26(69)&2(83)&
$-$4.3(12)&$-$0.6(6)&$-$3(4)\\[-0mm]
\hline\hline&&&&&&&&&&&&&&
\end{tabular}
\caption{\small Results from simultaneous fits to several observables based on
  ChPT at NLO and NNLO. Full expressions and the definitions of the
  coefficients $r_{V_1}, r_{V_2}, r_f$ and $r_M$ can be found in
  appendix~\ref{app:chiptB} }\label{tab:combinedfits}
\end{sidewaystable}
All our fit results are compiled in tables~\ref{tab:individualfits}
and~\ref{tab:combinedfits}. Fits to individual observables using ChPT at NLO
are listed in table~\ref{tab:individualfits}, while
table~\ref{tab:combinedfits} contains results of joint fits to more than one
observable, employing both NLO and NNLO expressions. First we note that
individual NLO fits to the pion decay constant and mass yield estimates for
$F$ as well as the LECs $\lc{3}$ and $\lc{4}$, which are in the same ballpark
than those given in the FLAG report (despite the fact that fitting the pion
mass with the decay constant fixed at its physical value gives relatively
large values of $\chi^2/\rm dof$). In particular, we find $r_0F=0.22-0.23$
which, using $r_0=0.503\,\fm$ from \cite{Fritzsch:2010aw}, translates into
$F=86-90$\,MeV. Given that the PDG quotes the physical pion decay constant as
$F_\pi^{\rm phys}=92.2$\,MeV, we find that the ratio $F_\pi/F$ determined in
this way is completely in line with the global FLAG estimate for this
quantity.

However, there is a significant downward shift of about 60\% for $r_0F$ when
the lattice data for the pion form factor or, alternatively, the charge radius
are fitted to the NLO formulae. This very low estimate for $F$ is accompanied
by a much smaller value for the LEC $\lc{6}$ compared to the range quoted in
the FLAG report. Taken at face value, such fits would suggest chiral
corrections as large as $60-70\%$ between the pseudoscalar decay constant in
the chiral limit and at the physical pion mass. Such a scenario contradicts
completely the experience gained in lattice calculations and effective field
theory analyses over many years. Therefore, despite the fact that NLO fits to
the form factor and the charge radius have good $\chi^2/\rm dof$, we conclude
that the results make no sense. If -- on the other hand -- one constrains
the pion decay constant to its physical value, one finds estimates
for $\lc{6}$ which are actually compatible with previous results. However,
such fits are not very plausible since their $\chi^2/\rm dof$ is unacceptably
large. 

The failure of the NLO formulae to describe the data for either the form
factor or the charge radius in a meaningful way is also manifest in the
modelling of cutoff effects of order~$a^2$. For instance, the value of the
coefficient $\alpha_f$ determined from a fit to the decay constant $F_\pi$
agrees with the result from a joint fit to both $F_\pi$ and $m_\pi^2$ (see the
corresponding entries marked ``NLO'' in tables~\ref{tab:individualfits}
and~\ref{tab:combinedfits}). When either $f_{\pi\pi}$ or $\langle
r_\pi^2\rangle$ is included into a joint NLO fit the coefficient $\alpha_f$
changes its sign and becomes positive. As can be clearly seen from
figure~\ref{fig:chipt-ext-B}, a positive value of $\alpha_f$ is incompatible
with the observation that $F_\pi$ increases as the lattice spacing is
reduced. 

To summarise: Chiral Perturbation Theory at NLO fails to produce a consistent
description of our lattice data for the entire set of pion observables within
the mass range considered in this paper. While individual and joint fits to
the pion mass and the pion decay constant lead to a coherent picture,
inconsistencies arise when comparing or combining the fits with the data for
the form factor or the charge radius.
\subsubsection{NNLO chiral fits}

\begin{figure}
\begin{minipage}[c]{.9\textwidth}
\centering
\noindent
 \includegraphics[width=\textwidth]{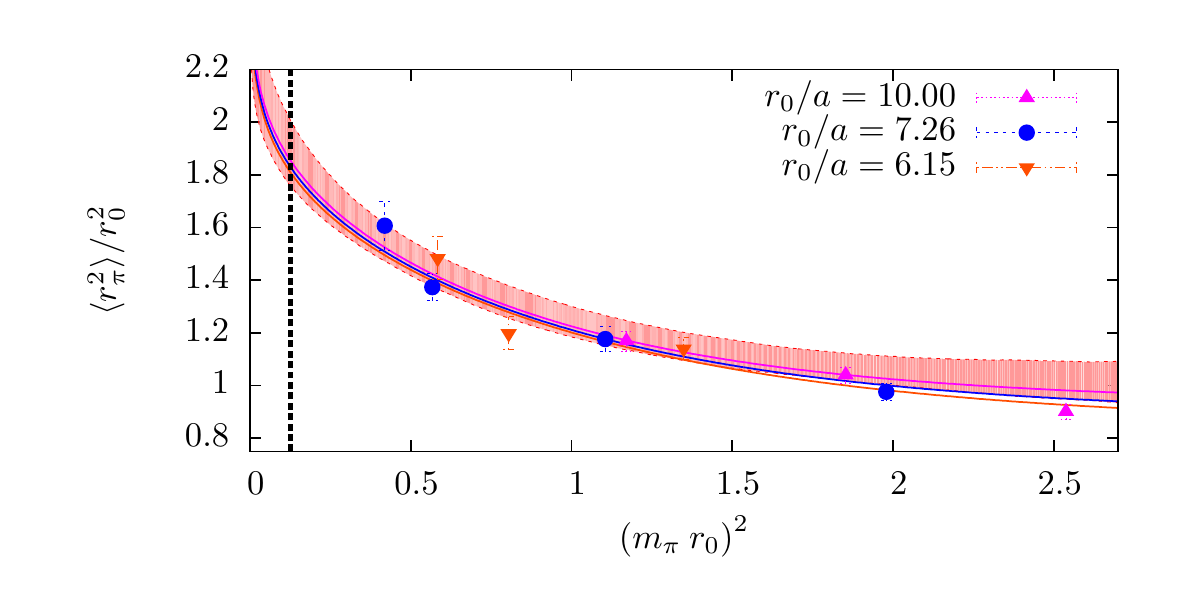}
\includegraphics[width=\textwidth]{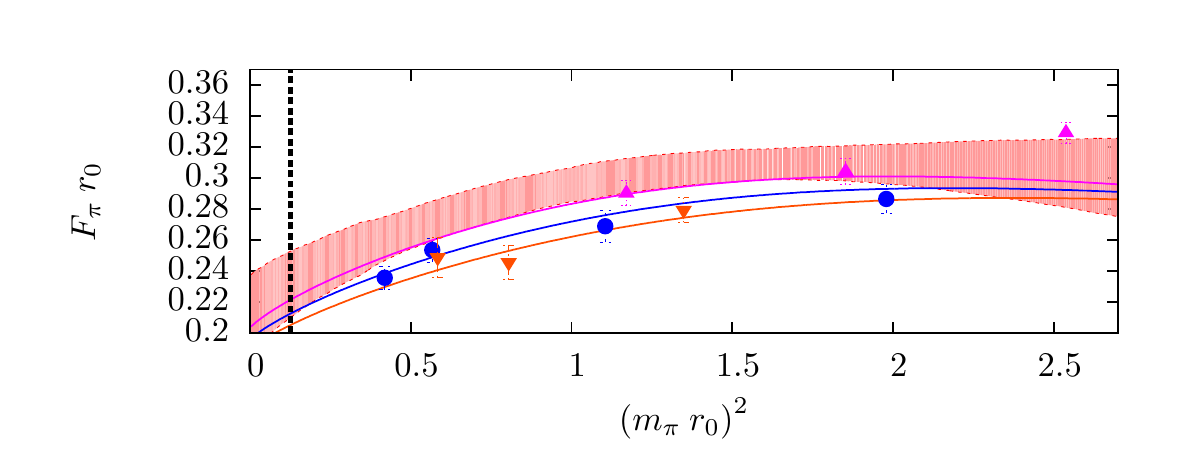}
\includegraphics[width=\textwidth]{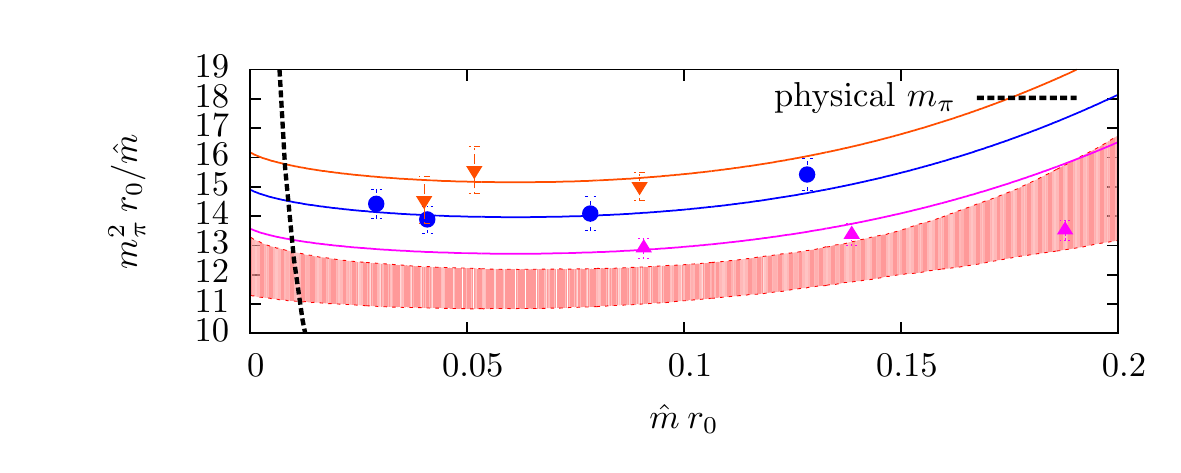}
\end{minipage}
\caption{\small Results from the global fit to the data of $f_{\pi\pi}$,
  $\Fpi$ and $m_\pi$ to ChPT at NNLO with a mass cut at 560~MeV. Shown are
  the chiral extrapolations for $\rpi$, $\Fpi$ and $m_\pi$ (from top to
  bottom). The red band represents the chiral behaviour of the quantity
  associated with the plot and the different solid lines are the results for
  the three different lattice spacings.}
\label{fig:chipt-ext-B}
\end{figure}

\begin{figure}
\begin{minipage}[c]{.9\textwidth}
\centering
\noindent
 \includegraphics[width=\textwidth]{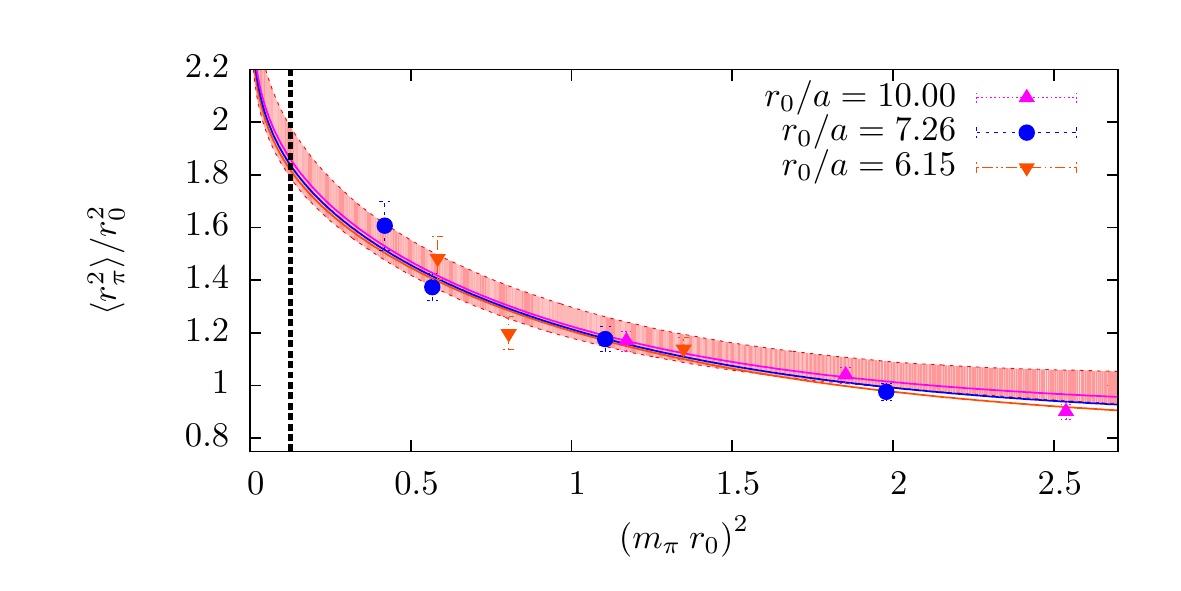}
 \newline
\includegraphics[width=\textwidth]{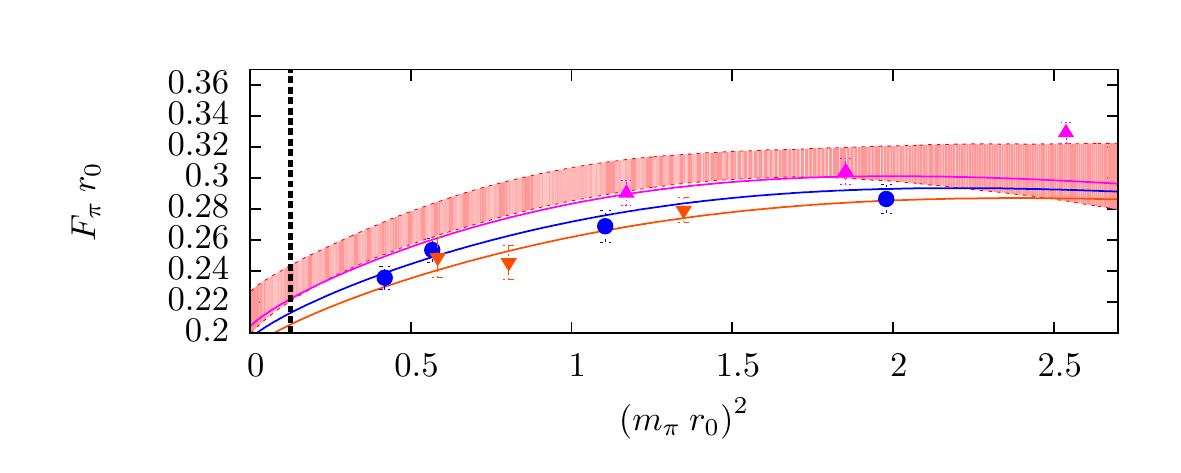}
 \newline
\includegraphics[width=\textwidth]{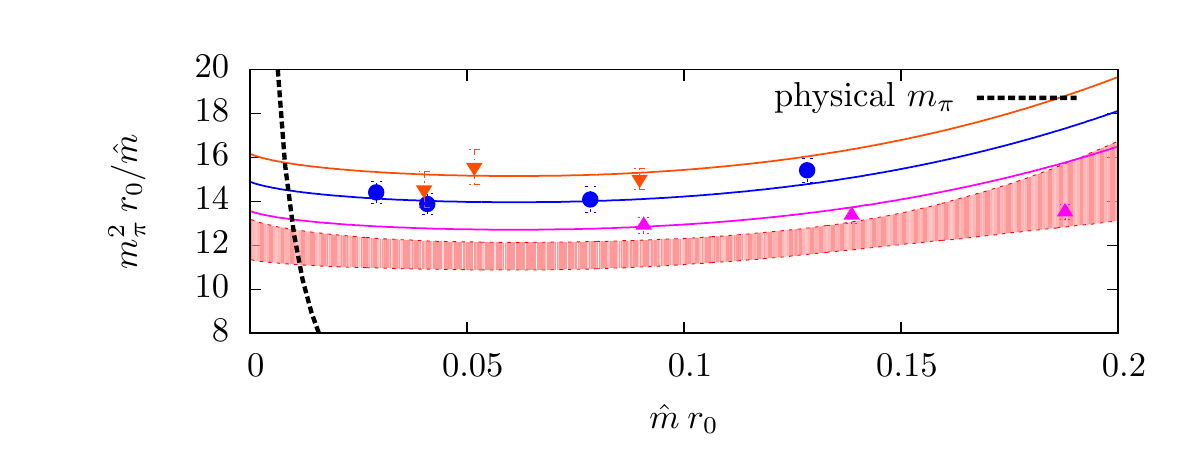}
\end{minipage}
\caption{\small Results from the global fit to the data of $\rpi$, $\Fpi$ and
  $m_\pi$ (from top to bottom) to ChPT at NNLO. The legend is
  the same as in figure~\ref{fig:chipt-ext-B}.}
\label{fig:chipt-ext-C}
\end{figure}

From the discussion above it is clear that a consistent description of our
data may be obtained either by extending the pion mass range to smaller values
or by going beyond NLO in ChPT. The NNLO expressions for $F_\pi$ and $m_\pi$
together contain eight LECs plus two parameters associated with cutoff
effects. After including $f_{\pi\pi}$ or $\langle r_\pi^2\rangle$ the number of
parameters increases to 14 and~13, respectively (cf.~\ref{app:chiptB}). We are
thus faced with the problem of having to constrain a large number of
parameters with a limited set of data points. It is then not surprising that
all our attempts at determining the full set of low-energy parameters were
unsuccessful. We note that similar difficulties were encountered by the ETM
Collaboration in their two-flavour study of the pion form
factor~\cite{Frezzotti:2008dr}. We therefore decided to stabilise the fits by
fixing two of the LECs, $\lc{1}$ and $\lc{2}$, to the values determined from
$\pi\pi$-scattering~\cite{Colangelo:2001df}, i.e.
\be
\label{eq:fix-l1-l2}
\lc{1}=-0.4(5) \quad \textnormal{and} \quad \lc{2}=4.3(1) \;.
\ee 
In the expressions for $m_\pi^2$, $F_\pi$ and $f_{\pi\pi}$ or $\rsq$ these
LECs appear only at NNLO. We checked explicitly that our results do not change
significantly when the central values of $\lc{1}$ and $\lc{2}$ are varied by
100\%. Moreover, the uncertainties for both LECs are fully included in the
analysis, by employing the same procedure described in
section~\ref{sec:fitting}. Even after reducing the number of free parameters
we found that only the joint fits to $(F_\pi, m_\pi^2)$ on the one hand and
either $(F_\pi, m_\pi^2, f_{\pi\pi})$ or $(F_\pi, m_\pi^2, \rsq)$ on the other
led to stable and consistent results, which are summarised in
table~\ref{tab:combinedfits}. Figures~\ref{fig:chipt-ext-B}
and~\ref{fig:chipt-ext-C} show the chiral extrapolations with $F_\pi$,
$m_\pi^2$ and $f_{\pi\pi}$ or $\rpi$, respectively, with a mass cut at
560\,MeV. The statistical uncertainty on the LECs increases noticeably as the
upper mass cut is lowered and less data points are allowed to constrain the
fit.

To summarise, the chiral expansion at NNLO provides a consistent
description of the data for $F_\pi$, $m_\pi^2$, $f_{\pi\pi}$ and $\rpi$. At
the current level of precision we do not observe severe inconsistencies like
in the case of NLO fits, and thus the results of the NNLO fits appear more
trustworthy.

\subsubsection{NLO and NNLO chiral fits: Final results}

In general, all fits based on the NNLO formulae are of reasonable quality in
terms of $\chi^2/\rm dof$. However, for a mass cut as low as 430~MeV the fit
ceases to be meaningful, as the central values become volatile while the
statistical errors increase significantly. We therefore decided to take the
simultaneous fit of $F_\pi$, $m_\pi^2$ and $\rpi$ at NNLO with an upper mass
cut of $m_\pi^{\rm cut}=560$~MeV as our reference result. Successful fits to
NLO expressions can only be achieved by excluding the data for the form factor
or charge radius, and we regard the joint fits to $F_\pi$ and $m_\pi^2$ for a
mass cut of 430\,MeV as our NLO reference results.  Table~\ref{tab:bestfit}
summarises our best fits based on both NLO and NNLO Chiral Pertubation Theory
expressions. The LECs extracted from either NLO or NNLO are compatible. Owing
to the better statistical accuracy, we take our final results for $F$, $B$,
$\lc{3}$ and $\lc{4}$ obtained from NLO ChPT as our best overall
estimates. The LEC $\lc{6}$ is extracted from the global NNLO fit.
\begin{table}
\begin{center}
\small       
\begin{tabular}{llccccccc}
\hline\hline\\[-3mm]
&&$m_\pi^{\rm cut}$&$\chi^2/\rm dof$&$r_0F$&$r_0B$&$\lc{3}$&$\lc{4}$&$\lc{6}$\\
\hline\\[-3mm]
NLO &$F_\pi$, $m_\pi^2$&430~MeV&0.8&0.213(17)&6.3(5) &3.0( 6)&4.7(3)\\
NNLO&$F_\pi$, $m_\pi^2$,
$\rpi$&560~MeV&1.3&0.213(14)&6.1(5)&4.0(13)&5.3(8)&15.5(15)\\
\hline\hline
\end{tabular}
\end{center}
\caption{\small Summary of best fits with NLO and NNLO chiral perturbation
  theory formulae.}\label{tab:bestfit}
\end{table}
\subsection{Fits guided by polynomial models}\label{sec:poly-ext}
\begin{figure}
\begin{center}
 \includegraphics[width=0.9\textwidth]{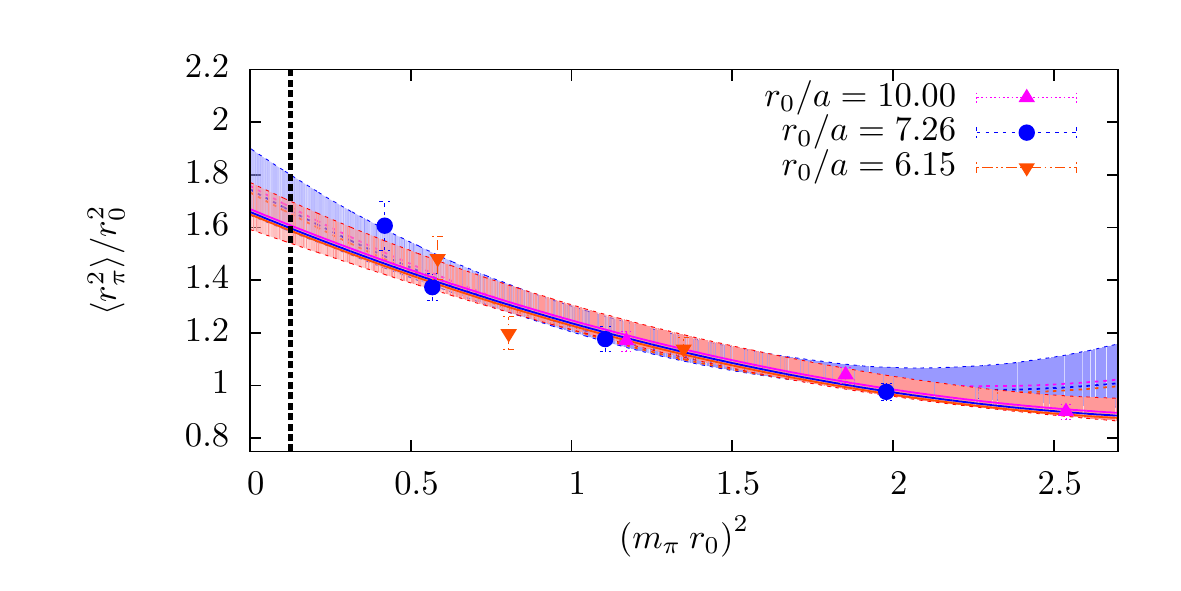}
\end{center}
\caption{\small Result for the fit to the form in eq.~\refeq{eq:poly-ch-extra}
  to order $m_\pi^4$, including lattice artefacts. The red band is the result
  for the fit in the continuum to all data points and the blue band is the
  result for the fit with a mass cut of 560\,MeV. The different colours
  emphasise different lattice spacings. The solid and dashed lines are the fit
  result for the different lattice spacings and all data points and the mass
  cut at 560~MeV, respectively. For the results at the different lattice
  spacings the error bars are left out for the purpose of visibility.}
\label{fig:poly-ch-extra}
\end{figure}

All fits to the lattice data carried out so far took advantage of a firm
theoretical prediction, based on chiral dynamics, for the dependence of
several observables on the pion mass and momentum transfer, in terms of a
common set of low-energy parameters. Nevertheless, it is interesting to study
the ability of simple fit {\it ans\"atze}, for which the expansion is not
constrained by symmetries, to describe the data. One such model is a simple
polynomial in the square of the pion mass. Clearly, in the absence of an
underlying dynamical theory which relates different observables, a global fit
to, say, $\rsq$, $\fpi$ and $m_\pi^2$ makes little sense. Here we only
consider $\rsq$ using the ansatz
\be
\label{eq:poly-ch-extra}
\frac{\rpi}{r_0^2} = b_0 +b_a\left(\frac{a}{r_0}\right)^2
  +b_1\:(r_0 m_\pi)^2 +b_2\:(r_0 m_\pi)^4 +\ldots \;,
\ee
where we have scaled all dimensionful quantities in units of $r_0$. The
results are summarised in table~\ref{tab:coeff-comp}. As indicated by the
value of $\chi^2/\rm dof$, the fits are of reasonable quality, despite the
fact that, contrary to the {\it ans\"atze} used in the previous sections, no
chiral logarithms are taken into account. We do, however, find an
unsatisfactory dependence of the extrapolated charge radius, when the mass cut
is lowered to 430\,MeV. This may be ascribed to the stronger sensitivity of
polynomial fits to fluctuations in the data near the physical pion mass. We
note that an additional term $b_3$ in the fit including a $m_\pi^6$-term is
not properly determined by our data set. Figure~\ref{fig:poly-ch-extra}
illustrates the extrapolations via polynomials for the cases of a polynomial
to $\mathcal{O}(m_\pi^4)$ with all data points and also for a mass cut imposed
at 560\,MeV. As can be seen, both extrapolations describe the data well and
yield compatible results for the whole range of pion masses.

\begin{table}
\begin{center}
\begin{tabular}{l|cccccccc}
\hline
\hline
Order & $m_\pi^{\rm cut}$ & $b_0$ & $b_a$ & $b_1$ & $b_2$&
$\chi^2/\rm dof$ \\
\hline
\hline
$m_\pi^4$ & all & 1.68 ( 9) & $-$1 (3) & $-$0.50 (10) & 0.08 ( 3) & 1.4 \\
 & 560\,MeV & 1.77 (13) & $-$2 (3) & $-$0.67 (19) & 0.15 ( 7) & 1.4 \\
 & 430\,MeV & 2.20 (27) & $-$2 (4) & $-$1.70 (57) & 0.72 (32) & 1.0 \\
\hline
$m_\pi^6$ & all & 1.92 (19) & $-$1 (3) & $-$1.1 (4) & 0.5 (3) & 1.1 \\
 & 560\,MeV & 2.44 (32) & $-$0 (3) & $-$2.8 (10) & 2.1 (9) & 0.6 \\
 & 430\,MeV & 3.08 (87) & 1 (5) & $-$5.3 (35) & 5.0 (41) & 0.9 \\
\hline
\hline
\end{tabular}
\end{center}
\caption{\small Results for the coefficients of the naive polynomial model for
  $\rpi$ as defined via eq.~\refeq{eq:poly-ch-extra}.}
\label{tab:coeff-comp}
\end{table}
\subsection{Results at the physical point}
We can now use the fit results from the previous section to determine the pion
charge radius and decay constant in the continuum limit and at the physical
pion mass. The latter is understood as the mass of the charged pion in QCD,
$(m_{\pi^+})^{\rm QCD}$, i.e. with the electromagnetic contributions
subtracted. Following the discussion in section~3.1 of the FLAG
report\,\cite{Colangelo:2010et}, we find that $(m_{\pi^+})^{\rm QCD}$ is, to a
good approximation, given by the physical mass of the neutral pion, i.e.
\be
(m_{\pi^+})^{\rm QCD} \simeq {m_{\pi^0}}=135\,\rm MeV.
\ee
The combined chiral and continuum fits yield the values of the fit parameters
in the continuum limit. Furthermore, we have corrected our input data for
finite-size effects, as described in section~\ref{sec:fvol}.

The results are summarised in table~\ref{tab:finalresults}, where we have only
included those fits, for which the ratio $F_\pi/F$ does not deviate from unity
by more than 10\% and which also have acceptable values of $\chi^2/\rm
dof$. We find that the estimates for the physical pion decay constant and
charge radius show practically no variation outside the quoted statistical
errors, with the possible exception of the charge radius extracted from
polynomial fits. As the mass cut is decreased to 430\,MeV, the statistical
accuracy of NNLO fits deteriorates.
\begin{table}
\begin{center}
\begin{tabular}{llllll}
\hline\hline\\[-3mm]
fit&&$m_\pi^{\rm cut}$&$F_\pi\:r_0$&$\langle r_\pi^2\rangle/r_0^2$\\[-0mm]
\hline&&&&&\\[-3mm]
NLO & $F_\pi$&all&0.241 (14)&\\
NLO &$F_\pi, m_\pi$&&0.237 (14)&\\
NNLO&$F_\pi, m_\pi$&&0.241 (13)&\\
NNLO&$F_\pi, m_\pi, \langle r_\pi^2\rangle$&&0.240 (11)&1.85 ( 9)\\
NNLO&$F_\pi, m_\pi, f_{\pi\pi}$&&0.248 (17)&1.82 (10)\\
poly&$\rsq, \mathcal{O}(m_\pi^4)$&&&1.62 ( 8)\\
\hline&&&&&\\[-3mm]
NLO & $F_\pi$&560~MeV&0.241 (15)&\\
NLO &$F_\pi, m_\pi$&&0.233 (16)&\\
NNLO&$F_\pi, m_\pi$&&0.233 (15)&\\
NNLO&$F_\pi, m_\pi, \langle r_\pi^2\rangle$&& 0.231 (12)&1.90 (10)\\
NNLO&$F_\pi, m_\pi, f_{\pi\pi}$&&0.231 (20)&1.90 (12)\\
poly&$\rsq, \mathcal{O}(m_\pi^4)$&&&1.70 (11)\\
\hline&&&&&\\[-3mm]
NLO & $F_\pi$&430~MeV&0.236 (15)&\\
NLO &$F_\pi, m_\pi$&&0.230 (16)&\\
NNLO&$F_\pi, m_\pi$&&0.235 (22)&\\
NNLO&$F_\pi, m_\pi, \langle r_\pi^2\rangle$&&0.219 (32)&1.89 (29)\\
NNLO&$F_\pi, m_\pi, f_{\pi\pi}$&&0.186 (20)&2.30 (34)\\
poly&$\rsq, \mathcal{O}(m_\pi^4)$&&&2.00 (21)\\
\hline&&&&&\\[-4mm]
\multicolumn{2}{l}{PDG}&&0.235( 4)& 1.79( 5)\\
\hline\hline
\end{tabular}
\end{center}
\caption{\small Results for the pion decay constant and charge radius at the
  physical point in units of $r_0$. The PDG values have been converted using
  $r_0=0.503(10)\,\fm$~\cite{Fritzsch:2012wq}.}
\label{tab:finalresults}
\end{table}

In the following we discuss the various sources of systematic error. We note
that the uncertainties of input parameters, such as renormalisation factors or
the estimates for the LECs $\lc{1}$ and $\lc{2}$ were folded into the
analysis. \\
\textit{Cutoff effects} -- Our data are compatible with {\it ans\"atze}
assuming a linear dependence on $a^2$, and at the current level of precision
we are not sensitive to higher-order lattice artefacts. In order to estimate
the size of residual discretisation errors of order~$a^4$ we make the
following exercise. First, we note that the global fit to $m_\pi^2$, $F_\pi$
and $\rpi$ with $m_\pi^{\rm cut}=560$\,MeV suggests that the corrections of
order~$a^2$, estimated as the difference between the continuum limit and the
coarsest lattice spacing, amount to 10\% for the pion decay constant and 7\%
for the charge radius. This is in the same ballpark than the crude estimate of
$\rmO(a^2)$ lattice artefacts of $(\Lambda_{\rm QCD}a)^2\approx4\%$ (where we
have used $\Lambda_{\rm QCD}\approx500\,\MeV$). By the same argument one can
give a rough estimate of $\rmO(a^4)$ cutoff effects, which then amounts to
0.2\%. Compared to the typical statistical accuracy this error is
negligible.\\
\textit{Finite size effects} -- All our ensembles satisfy $m_\pi L > 4$, which
has often been considered sufficient to guarantee small effects due to the
finiteness of the box size. In addition, we have corrected for finite-volume
effects on the pion mass and decay constant using ChPT at NNLO and ChPT at NLO
for the form factor and charge radius. We believe that the residual finite
volume effects are negligible.\\
\textit{Chiral extrapolation} -- The only globally consistent extrapolation of
the lattice data was achieved using NNLO chiral perturbation theory, while NLO
turned out to be sufficient when fitting only $F_\pi$ and $m_\pi$. We
estimated the residual uncertainty due to chiral extrapolation from the spread
of results obtained considering different mass cuts. For the results covered
by the NLO extrapolation (i.e. $F$, $F_\pi$, $\lc{3}$, $\lc{4}$ and $B$) we
used the difference between $m_\pi^{\rm cut}=430$\,MeV and 560\,MeV as the
symmetric systematic error due to the chiral extrapolation. For $\rsq$ and
$\lc{6}$ determined via the NNLO fit we use the spread between the central
value between the fit over all data and the one with an upper mass cutoff of
560\,MeV.\\
\textit{Scale setting} -- The Sommer scale~\cite{Sommer:1993ce} was used to
combine data obtained at different values of the lattice spacing and to convert
to physical units. The absolute physical scale was set by the kaon leptonic
decay constant~\cite{Fritzsch:2012wq}, and the result for the scale is fully
compatible with the independent determination using the mass of the Omega
baryon~\cite{Capitani:2011fg}. The associated errors were folded into the
analysis during the resampling (c.f. section~\ref{sec:fitting}).\\
{
\textit{Critical slowing down}
-- It has been known for some time that simulations of lattice QCD
suffer from critical slowing down, which rapidly accelerates
when approaching the 
continuum limit~\cite{DelDebbio:2002xa,Schaefer:2009xx,Schaefer:2010hu}.
While promising ideas for reducing the severity of the problem in future
simulations have by now been developed~\cite{Luscher:2011kk,Luscher:2012av},
we cannot exclude the possibility that the results of this paper which
are based on the ensembles at our finest lattice spacing (N3, N4 and N5)
are affected. One consequence for the data
analysis, the underestimation of autocorrelations and hence the underestimation
of statistical errors, was studied 
in~\cite{Schaefer:2010hu} where a procedure for estimating this effect
 has been devised. The limited number of measurements in the present
 work, however, do not allow for a similar treatment of the observables 
 considered here. In this situation we have mimicked the effect
 of an underestimation of the statistical error in all results generated from
 ensembles N3, N4 and N5, by inflating the statistical error by a factor of two
 prior to all subsequent analysis. The size of the inflated error is 
 suggested by the findings
 of~\cite{Schaefer:2010hu,Fritzsch:2012wq}. In general, the central values of 
 the final results hardly change, and only the statistical error increases 
 slightly. In the following we adopt this statistical error when quoting
 the final results but keep the central values from the analysis without
 the inflated error.
}

We now summarise our final results. Since the combined NLO fits to the pion
decay constant and the pion mass are of good quality and statistical accuracy,
we use it to quote final results for the bulk of the fitted low-energy parameters,
imposing a mass cut of 430\,MeV:
\begin{equation}
\begin{array}{lllllll}
F_\pi   &=&90(8)(2)\,{\rm MeV}, &F  &=&84(8)(2)\,{\rm MeV},\\
F_\pi/F &=&1.080(16)(6),        &B  &=&2.5(3)(1)\,{\rm GeV},\\
\lc{3}	&=&3.0(7)(5),           &\lc{4}&=&4.7(4)(1).
\end{array}
\end{equation}
Note that the low energy constant $B$ depends on the renormalisation scheme.
Here we quote the result in the $\overline{\rm MS}$-scheme at
$\mu=2$~GeV. Since the product of $B$ and the current (PCAC) quark mass is
scale and scheme independent, the LEC $B$ in the $\msbar$-scheme is obtained
after dividing by the renormalisation factor of the quark mass. For our chosen
discretisation, this factor (0.968(20)) 
is easily determined using the results of
refs.~\cite{DellaMorte:2005kg,Fritzsch:2012wq}.

For the pion charge radius and the LEC $\lc{6}$ we take the results from the
combined NNLO-fit to the pion decay constant, the pion mass and the pion
charge radius,
\begin{equation}
\begin{array}{lllllll}
\rpi   &=&0.481(34)(13){\rm fm}^2,\\
\lc{6} &=&15.5(1.7)(1.3).\\
\end{array}
\end{equation}
In each case the first error is statistical and the second one is due to the
chiral extrapolation, as explained above. Other systematic effects are much
smaller and have therefore not been specified.

We end this section with the observation that our results for $F_\pi/F,
\lc{3}$ and $\lc{4}$ are in very good agreement with the values listed in
section~4 of the FLAG report~\cite{Colangelo:2010et}. Furthermore, we can use
the well-known Gell-Mann-Oakes-Renner (GMOR) relation~\cite{GellMann:1968rz} 
\be
  \Sigma = F^2 \; B \;, 
\ee
which relates the LECs $B$ and $F$ to the quark condensate $\Sigma$. Our
result for the condensate in the $\overline{\rm MS}$-scheme at a
renormalisation scale of $\mu=2$~GeV is
\be
 \Sigma^{1/3} = 261 \:(13)(1) \:{\rm MeV} \;, 
\ee
which is also in good agreement with the results listed
in~\cite{Colangelo:2010et}.


\section{Conclusions and outlook}\label{sec:conclusions}
\begin{table}[t]
\footnotesize
\begin{center}
\begin{tabular}{l@{\hspace{0mm}}c@{\hspace{1mm}}c|l|l|c|l}
\hline
\hline&&&\\[-3mm]
 & ref & $N_f$& chiral extrapolation & $\rpi$ extr. &
 $\rpi/r_0^2$ &\multicolumn{1}{c}{ $\rpi$~[fm$^2$]} \\
\hline
\hline
this study  &&2& NNLO ChPT ($\rpi,\;\Fpi,\;m_\pi$) & poly. $\Ord{q^4}$ &
   1.90(11) & 0.481(33)(13) \\
\hline
QCDSF & \cite{Brommel:2006ww} &2& poly. for $M_{\rm pole}$ & pole &
   2.027(89) & 0.442(19) \\
ETMC  & \cite{Frezzotti:2008dr} &2& NNLO ChPT ($\rpi,\;\Fpi,\;m_\pi$) &
  pole & & 0.456(30)(24) \\
JLQCD/TWQCD & \cite{Aoki:2009qn} &2& NNLO ChPT ($\rpi,\;\Fpi,\;m_\pi$) &
  pole & 1.703(96) & 0.409(23)(37) \\
RBC/UKQCD & \cite{Boyle:2008yd}&2+1 & NLO ChPT ($\fpipi$) & NLO & &
  0.418(31) \\ 
Nguyen et al. & \cite{Nguyen:2011ek} &2+1 & NNLO ChPT
($\rpi,\;\Fpi,\;m_\pi$) & pole & & 0.441(46) \\
\hline
PDG & \cite{Nakamura:2010zzi} & --- & --- & gl. av. & & 0.452(11) \\
Amendolia et al. & \cite{Amendolia:1986wj} & ---& --- & pole & & 0.439(8) \\
BCT & \cite{Bijnens:1998fm} & --- & --- & NNLO & & 0.437(16) \\
\hline
\hline
\end{tabular}
\caption{\small Compilation of results for the charge radius at the physical
  point from lattice
  QCD~\cite{Brommel:2006ww,Frezzotti:2008dr,Aoki:2009qn,Boyle:2007wg,
    Boyle:2008yd,Nguyen:2011ek}, including this study,
  experiment~\cite{Nakamura:2010zzi,Amendolia:1986wj} and ChPT to NNLO for
  the experimental data~\cite{Bijnens:1998fm}. See also scatter plot in 
  figure~\ref{fig:comparison}.
  }
\label{tab_fpp-res_disc.1}
\end{center}
\end{table}
\begin{figure}
 \begin{center}
 \includegraphics[width=12cm]{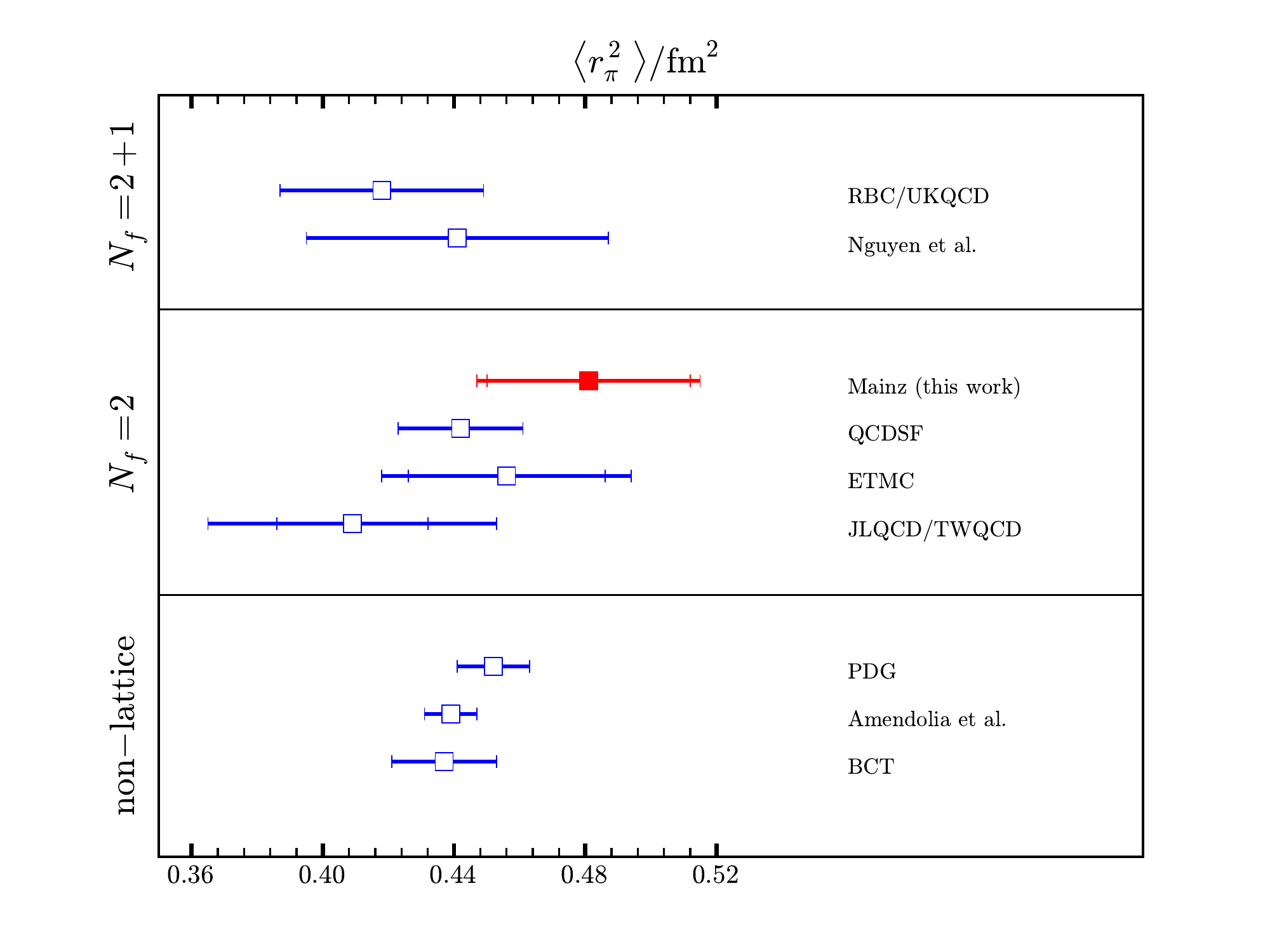}
 \end{center}
 \caption{Comparison of results for the charge radius, cf. 
 table~\ref{tab_fpp-res_disc.1}.}\label{fig:comparison}
\end{figure}

We have presented the first determination of the iso-vector electromagnetic
form factor which does not rely on any particular model for the
$Q^2$-dependence of the form factor. Our study in two-flavour QCD has full
control over the main systematic uncertainties, except isospin breaking
effects. A crucial ingredient was the extensive use of partially twisted
boundary conditions, which allowed us to achieve a high resolution of data
points for the form factor close to $Q^2=0$, thereby enabling a
model-independent determination of the charge radius.

Our data for the pion mass, decay constant, form factor and charge radius were
then subjected to extensive fits to ChPT at NLO and NNLO, augmented by terms
which parameterise leading lattice artefacts. 
{While the NLO expressions failed to produce a consistent
description of all observables, individual or joint fits to the data of the
decay constant and mass lead to a coherent picture. This indicates 
a problem with the effective theory description of the form factor at NLO
at least for the range of quark masses conisdered here. 
At the level of statistical
precision of the lattice data achieved here the NNLO expressions on the other 
hand allow
for a fully consistent description of all observables.}
The {proliferation} of free parameters at NNLO could
be dealt with by fixing two LECs, $\lc{1}$ and $\lc{2}$, from $\pi\pi$
scattering.

The ability of ChPT to describe lattice data generally depends on the mass
range considered in simulations. We note that our conclusions have been
reached for pion masses between 280 and 540~MeV. It will be interesting to
study whether ChPT at NLO can be successfully fitted to the data including the
form factor or the charge radius, when data at or near the physical pion mass
become available.

In table~\ref{tab_fpp-res_disc.1} and figure~\ref{fig:comparison}
we compare our result for the charge radius
to those from other lattice simulations, experimental determinations, as well
as results from a ChPT description of experimental data. Despite the
relatively good agreement between the various lattice estimates and the value
quoted by the PDG\,\cite{Nakamura:2010zzi}, one observes a certain spread
among the lattice results, with the result of this study at the upper end. In
some cases the differences can be traced to the scale setting
procedure.\footnote{For instance, the QCDSF Collaboration quotes
  $r_0=0.467\,\fm$, while we take the more recent determination of
  ref.\,\cite{Fritzsch:2012wq}, i.e. $r_0=0.503(10)\,\fm$} It is then clear
that further efforts in lattice QCD are required to pin down the pion charge
radius with better overall accuracy. The need for additional simulations --
preferably at the physical pion mass -- is further highlighted by the
difficulties which we encountered in obtaining a consistent ChPT description
of the data for the form factor and charge radius.\\[3mm]
{\bf Acknowledgements:}
We are grateful to our colleagues within the CLS project for sharing gauge
ensembles. Calculations of correlation functions were performed on the
dedicated QCD platform ``Wilson'' at the Institute for Nuclear Physics,
University of Mainz and on the QCD HPC ``thqcd2'' cluster at CERN.  We thank
Dalibor Djukanovic for technical support.  This work was supported by DFG
(SFB443 and SFB TR55 ``Hadron Physics from Lattice QCD'') 
and the Research Center EMG funded by Forschungsinitiative
Rheinland-Pfalz.  The research leading to these results has also received
funding from the European Research Council under the European Community's
Seventh Framework Programme (FP7/2007-2013) ERC grant agreement No 279757.

\begin{appendix}
\section{Appendix}
\subsection{Chiral perturbation theory to NNLO}\label{app:chiptA}
For the purpose of performing a global fit to the data of $\mpi,\:\fpi$ and
$\fpipi$ (or $\rpi$ alternatively) we first review the formulae of chiral
perturbation theory to NNLO as given in~\cite{Bijnens:1998fm}. For
convenience we adopt their notation and define the quantities
\be
\label{eq:A1}
\begin{array}{rclcrclcrcl}
\xt & \equiv & \frac{\mpi^2}{\fpi^2} & \qquad &
m_0 & \equiv & 2\:B\:\hat{m} & \qquad &
\bq & \equiv & \frac{q^2}{\mpi^2} \\
N & \equiv & 16\:\pi^2 & &
L & \equiv & \displaystyle \frac{1}{N} \: \ln \left(\frac{\mpi^2}{\mu^2}\right)
& & &
\end{array}
\ee
Here $\mu$ is the renormalisation scale which we set to
$\mu=m_\rho^{phys}=0.77$~GeV~\cite{Nakamura:2010zzi} and $\hat{m}$ is the
renormalised bare quark mass. Note that a different
renormalisation of $\hat{m}$ is absorbed in a different renormalisation of the
LEC $B$. The LECs that are scale-independent are
denoted as $\lc{i}$ and are given in terms of the renormalised LECs at the
physical pion mass. For these we define the related scale-dependent quantities
\be
\label{eq:A2}
\lcr{i} \equiv \frac{\gamma_i}{2\:N} \:
\left(\lc{i}+N\:\left.L\right|_{m_\pi^{\rm phys}}\right) \quad
\text{and} \quad k_i \equiv \left( 4\:\lcr{i} - \gamma_i \: L \right) \: L
\ee
which appear in the formulae. Here $\gamma_i$ are the anomalous dimensions,
given by
\be
\label{eq:A3}
\gamma_1 = 1/3 \:, \quad \gamma_2 = 2/3 \:, \quad \gamma_3 = - 1/2 \:, \quad
\gamma_4 = 2 \:, \quad \gamma_6 = - 1/3 \;,
\ee
and $\left.L\right|_{m_\pi^{\rm phys}}$ denotes the chiral logarithm with the
physical pion mass in the numerator.
We further define the functions
\be
\label{eq:A4}
\begin{array}{rcl}
 J(\bq) & \equiv & \displaystyle \frac{\sqrt{z}}{N} \: \ln \left(
\frac{\sqrt{z}-1}{\sqrt{z}+1} \right) + \frac{2}{N} \vspace*{2mm} \\
 K_1(\bq) & \equiv & \displaystyle z\:h^2(\bq) \vspace*{2mm} \\
 K_2(\bq) & \equiv & \displaystyle z^2\:h^2(\bq)-\frac{4}{N^2} \vspace*{2mm} \\
 K_3(\bq) & \equiv & \displaystyle \frac{N\:z}{\bq} \: h^3(\bq) +
\frac{\pi^2\:h(\bq)}{N\:\bq} - \frac{\pi^2}{2\:N^2} \vspace*{2mm} \\
 K_4(\bq) & \equiv & \displaystyle \frac{1}{\bq\:z} \: \left( \frac{1}{2} \:
K_1(\bq) + \frac{1}{3} \: K_3(\bq) + \frac{1}{N} \: J(\bq) +
\frac{(\pi^2-6)\:\bq}{12\:N^2} \right) \;,
\end{array}
\ee
with
\be
\label{eq:A5}
 z \equiv 1 - \frac{4}{\bq} \quad \text{and} \quad h(\bq) \equiv
\frac{1}{N\:\sqrt{z}} \: \ln \left( \frac{\sqrt{z}-1}{\sqrt{z}+1} \right)
\ee

Using these quantities the pion mass and the pion decay constant to NNLO are
given by
\be
\label{eq:A6}
\mpi^2 = m_0 \: \left\{ 1 + \xt \: \mnlo + \xt^2 \: \mnnlo \right\} \quad
\text{and} \quad \fpi = F \: \left\{ 1 + \xt \: \fnlo + \xt^2 \: \fnnlo
\right\}
\ee
where
\be
\label{eq:A7}
\begin{array}{rcl}
\displaystyle \mnlo & \equiv & \displaystyle 2\:\lcr{3} + \frac{1}{2}\:L
\vspace*{2mm} \\
\displaystyle \mnnlo & \equiv & \displaystyle \frac{1}{N} \: \left( \lcr{1} +
2\:\lcr{2} - \frac{13}{3}\:L \right) + \frac{163}{96\:N^2} - \frac{7}{2} \: k_1
- 2\:k_2 - 4\:(\lcr{3})^2 + 4\:\lcr{3}\:\lcr{4} \vspace*{2mm} \\
 & - & \displaystyle \frac{9}{4}\:k_3 + \frac{1}{4}\:k_4 + \rmr \vspace*{2mm} \\
\displaystyle \fnlo & \equiv & \displaystyle \lcr{4}-L \vspace*{2mm} \\
\displaystyle \fnnlo & \equiv &  \frac{1}{N} \: \left( -\frac{1}{2}\:\lcr{1} -
\lcr{2} + \frac{29}{12} \: L \right) - \frac{13}{192\:N^2} + \frac{7}{4}\:k_1 +
k_2 - 2\:\lcr{3}\:\lcr{4} \vspace*{2mm} \\
 & + & \displaystyle 2\:(\lcr{4})^2 - \frac{5}{4}\:k_4 + \rfr
\end{array}
\ee
Here $\rmr$ and $\rfr$ are constants stemming from the $\Ord{p^6}$ Lagrangian
after minimal subtraction and $F$ is the pion decay constant in the chiral
limit. For the form factor the expansion is written as
\be
\label{eq:A8}
\fpp(\bq) = 1 + \xt \: \ffnlo + \xt^2 \: \left( P_V^{(2)} + U_V^{(2)} \right)
\;.
\ee
Here $\ffnlo$ is given by
\be
\label{eq:A9}
\ffnlo \equiv \frac{1}{6} \: \left(\bq-4\right) \: J(\bq)
+ \bq \left( -\lcr{6}-\frac{1}{6}\:L-\frac{1}{18\:N} \right)
\ee
and $P_V^{(2)}$ and $U_V^{(2)}$ are the polynomial and dispersive NNLO piece
respectively, given by
\be
\label{eq:A10}
\begin{array}{rcl}
\displaystyle P_V^{(2)} & = & \displaystyle \bqf \: \left[ \frac{1}{12}\:k_1 -
\frac{1}{24}\:k_2 + \frac{1}{24}\:k_6 \right. \vspace*{2mm} \\
 & & \displaystyle \quad \quad \left. + \frac{1}{9\:N}
\:\left(\lcr{1}-\frac{1}{2}\:\lcr{2}+\frac{1}{2}\lcr{6}-\frac{1}{12}\:L-\frac{1}
{384}-\frac{47}{192\:N}\right) + \rvrt \right] \vspace*{2mm} \\
 & + & \displaystyle \bq \: \left[ -\frac{1}{2}\:k_1 + \frac{1}{4}\:k_2 -
\frac{1}{12}\:k_4 +
\frac{1}{2}\:k_6 - \lcr{4}\:\left(2\:\lcr{6}+\frac{1}{9\:N}\right) \right.
\vspace*{2mm} \\
 & & \displaystyle \quad \quad \left. + \frac{23}{36}\:\frac{L}{N} +
\frac{5}{576\:N} + \frac{37}{864\:N^2} + \rvro \right] \vspace*{2mm} \\
\displaystyle U_V^{(2)} & = & \displaystyle J(\bq) \: \left[
\frac{1}{3}\:\lcr{1}\:\left(-\bqf+4\:\bq\right) +
\frac{1}{6}\:\lcr{2}\:\left(\bqf-4\:\bq\right) +
\frac{1}{3}\:\lcr{4}\:\left(\bq-4\right) \right. \vspace*{2mm} \\
 & & \displaystyle \quad \quad \quad
+\frac{1}{6}\:\lcr{6}\:\left(-\bqf+4\:\bq\right) -
\frac{1}{36}\:L\:\left(\bqf+8\:\bq-48\right) \vspace*{2mm} \\
 & & \displaystyle \quad \quad \quad \left. + \frac{1}{N} \:
\left(\frac{7}{108}\:\bqf-\frac{97}{108}\:\bq+\frac{3}{4}\right) \right] +
\frac{1}{9}\:K_1(\bq) \vspace*{2mm} \\
 & + & \displaystyle \frac{1}{9}\:K_2(\bq)\:\left(\frac{1}{8}\:\bqf-\bq+4\right)
+ \frac{1}{6}\:K_3(\bq)\:\left(\bq-\frac{1}{3}\right) - \frac{5}{3}\:K_4(\bq)
\;.
\end{array}
\ee
$\rvro$ and $\rvrt$ are again coming from the $\Ord{p^6}$ Lagrangian.
\subsection{Reformulation of ChPT to NNLO for global fits}\label{app:chiptB}
For the intended fits as discussed in section \ref{sec:extrapolations} it is
necessary to
reorganize the chiral expansion, since the right hand sides depend
on $\mpi$ and $\fpi$ themselves. In
this appendix we describe the necessary replacements and list the results.
\subsubsection{Conventions and necessary replacements}
For the fits including $\fpipi$ it is convenient to define the new fit
parameter
\be
\label{eq:B1}
\ltil \equiv \lcr{1}-\frac{1}{2}\:\lcr{2}
\ee
which replaces the fit parameters $\lcr{1}$ and $\lcr{2}$ completely in
eq.~\refeq{eq:A10}. We also use $\ltil$ instead $\lcr{2}$ in $\mpi^2$ and $\fpi$
to have a consistent set of fit parameters.

To make the fit formulae self-consistent we have to replace $\mpi$ and $\fpi$
in each formula with the expressions in eq.~\refeq{eq:A6} and keep all terms to
$\Ord{\xt^2}$. In practice this means we have to replace $\mpi$ and $\fpi$
with its NLO
expressions in each NLO term. In the course of this replacement the parameters
\be
\label{eq:B2}
\xt\;,\quad L\;,\quad \bq\;, \quad \text{and} \quad J(\bq)
\ee
are modified. In the results similar parameters appear with the first order
parameters $m_0$
and $F$ instead of $\mpi^2$ and $\fpi$. We thus define:
\be
\label{eq:B3}
\begin{array}{rclcrcl}
\hxt & \equiv & \displaystyle \frac{m_0}{F^2}\;; & \quad &
\hbq & \equiv & \displaystyle \frac{q^2}{m_0}\;; \vspace*{2mm} \\
\hat{L}\;; & \equiv &
\displaystyle \frac{1}{N} \: \ln\left(\frac{m_0}{\mu^2}\right)\;; & &
\hlcr{i} & \equiv & \displaystyle \frac{\gamma_i}{2\:N} \:
\left(\lc{i}+N\:\hat{L}\right)\;; \vspace*{2mm} \\
 \hat{k}_i & \equiv & \displaystyle \left( 4\:\hlcr{i} - \gamma_i \: \hat{L}
\right) \: \hat{L}\;; & & \hat{z} & \equiv & 1 - \frac{4}{\hbq} \;.
\end{array}
\ee
As shorthand notation we further define
\be
\label{eq:B4}
\Delta_m \equiv 2\:\hlcr{3}+\frac{1}{2}\:\hat{L} \quad \text{and} \quad \Delta_f
\equiv \hlcr{4}-\hat{L} \;.
\ee
This is convenient, since in the following one just has to set $\Delta_m$ and
$\Delta_f$ to zero to obtain the NNLO formulae of the previous appendix.

Most of the replacements are straight-forward. The only more complicated
replacement is the one for the function $J(\bq)$, since it is a non-trivial
function of $z$. We write the result as
\be
\label{eq:B5}
J(\bq) = J(\hbq) - \hxt\:\frac{\Delta_m}{\hbq-4}\:\left( 2 \: J(\hbq) -
\frac{\hbq}{N} \right)\;.
\ee
%
\subsubsection{Reformulated formulae}
We now list the reformulated formulae for $\mpi^2$, $\fpi$ and $\fpipi$. The
pion mass is given by
\be
\label{eq:B7}
\mpi^2 = m_0 \: \left\{ 1 + \hxt \: \mnlo + \hxt^2 \: \left( \mnnlo + \Delta_m
\:
\left[ \Delta_m - 2 \: \Delta_f \right] + \frac{\Delta_m}{2\:N} \right) \right\}
\;,
\ee
where $\mnlo$ and $\mnnlo$ are defined as in eq.~\refeq{eq:A7} with every
quantity replaced by its modified version from eq.~\refeq{eq:B3}. Similarly the
modified ChPT expression for the pion decay constant is
\be
\label{eq:B8}
\fpi = F \: \left\{ 1 + \hxt \: \fnlo + \hxt^2 \: \left( \fnnlo + \Delta_f \:
\left[ \Delta_m - 2 \: \Delta_f \right] - \frac{\Delta_m}{N} \right) \right\}
\;,
\ee
where again the replacements in $\fnlo$ and $\fnnlo$ are implied. For
$\fpp(\hbq)$ we obtain
\be
\label{eq:B9}
\begin{array}{rcl}
\fpp(\bq) & = & 1 + \hxt \: \ffnlo + \hxt^2 \: \left( P_V^{(2)} + U_V^{(2)}
\right) \vspace*{2mm} \\
 & + & \hxt^2 \: \left[ -\Delta_m\:J(\hbq) - 2\:\Delta_f \:
\left( \frac{1}{6}\:\left(\hbq-4\right)\:J(\hbq)
- \hbq\:\left[\hlcr{6}+\frac{1}{6}\hat{L}+\frac{1}{18\:N}\right] \right)
\right]
\end{array}
\ee
with implied replacements in $\ffnlo$, $P_V^{(2)}$ and $U_V^{(2)}$.

For the pion charge radius as defined in eq.~\refeq{eq:chrad} the results
above yield
\be
\label{eq:B10}
\rpi = \frac{1}{m_0} \: \left\{ \hxt \: \left[\rpi\right]_1 + \hxt^2 \: \left(
\left[\rpi\right]_2 - 2\:\Delta_F \: \left[\rpi\right]_1 -
\frac{\Delta_m}{N} \right) \right\}
\ee
where $\left[\rpi\right]_1$ is the NLO part as given in \cite{Bijnens:1998fm},
\be
\label{eq:B11}
\left[\rpi\right]_1 = - \left(6\:\hlcr{6}+\hat{L}+\frac{1}{N}\right) \;,
\ee
and $\left[\rpi\right]_2$ the usual NNLO part,
\be
\label{eq:B12}
\left[\rpi\right]_2 = - 12\:\hat{L}\:\ltil - \frac{1}{2}\:k_4 +
3\:k_6 - 12\:\hlcr{4}\:\hlcr{6} + \frac{1}{N} \: \left( - 2\:\hlcr{4} +
\frac{31}{6}\:\hat{L} + \frac{13}{192} - \frac{181}{48\:N} \right) + 6\:\rvro
\;.
\ee
Note, that the results discussed above are in agreement with the ones listed
in~\cite{Frezzotti:2008dr}.
\subsubsection{Inclusion of lattice artefacts}
Since the data indicates the presence of residual lattice artefacts it is
desirable to include these effects in the chiral extrapolation. To this end we
extend the formulae from the last section of the appendix used for the global
fits to the more general form:
\begin{eqnarray}
 \label{eq:B13}
\left(\mpi\:\left[1+\alpha_m\frac{a^2}{r_0^2}\right]\right)^2 & = &
\refeq{eq:B7} \;; \\
 \label{eq:B14}
\fpi & = & \refeq{eq:B8} + \alpha_f\frac{a^2}{r_0^2} \;; \\
 \label{eq:B15}
\fpp(\hbq) & = & \refeq{eq:B9} + \alpha_r\frac{a^2}{r_0^2} \:
\frac{m_0\:\hbq}{6} \;; \\
 \label{eq:B16}
\rpi & = & \refeq{eq:B10} + \alpha_r\frac{a^2}{r_0^2} \;.
\end{eqnarray}
Note that for $m_\pi$ the lattice artefacts are expected to be of $\Ord{a^2}$
for the mass itself and thus should be included on the left hand side.

\end{appendix}

\vspace{.8cm}

\bibliographystyle{JHEP}
\bibliography{spires}

\end{document}